\documentclass[preprint,12pt,authoryear]{elsarticle}

\usepackage{amssymb}
\usepackage{amsmath}
\usepackage{booktabs} 
\usepackage{hyperref}
\usepackage{adjustbox}
\usepackage{lineno}

\journal{Journal of Wind Engineering $\&$ Industrial Aerodynamics }
\begin{document}

\begin{frontmatter}



\title{Stable and convective boundary-layer flows in an urban array}


\author[author]{Davide Marucci}
\author[author]{Matteo Carpentieri\corref{mycorrespondingauthor}}
\ead{m.carpentieri@surrey.ac.uk}
\address[author]{EnFlo Laboratory, Department of Mechanical Engineering Sciences, University of Surrey, Guildford, Surrey GU2 7XH, UK}
\cortext[mycorrespondingauthor]{Corresponding author}

\begin{abstract}
In this paper non-neutral approaching flows were employed in a meteorological wind tunnel on a regular urban-like array of rectangular buildings. As far as stable stratification is concerned, results on the flow above and inside the canopy show a clear reduction of the Reynolds stresses and an increment of the Monin-Obukhov length up to 80\%. The roughness length and displacement height were also affected, with a reduction up to 27\% for the former and an increment up to 5\% for the latter. A clear reduction of the turbulence within the canopy was observed.
In the convective stratification cases, the friction velocity appears increased by both the effect of roughness and unstable stratification. The increased roughness causes a reduction in the surface stratification, reflected in an increase of the Monin-Obukhov length, which is double over the array compared to the approaching flow. The effect on the aerodynamic roughness length and displacement height are specular to the SBL case, an increase up to 55\% of the former and a reduction of the same amount for the latter.
\end{abstract}



\begin{keyword}
Stable boundary layer \sep
Convective boundary layer \sep
Wind tunnel \sep
Array of cuboids \sep
Urban flows


\end{keyword}

\end{frontmatter}


\section{Introduction}
Understanding flow and gas dispersion in urban areas is becoming increasingly important due to rapid urbanisation and related health and economic issues. One aspect of urban flow and dispersion that is still not very well studied is atmospheric stratification, despite the fact that cities very often present either stable or convective conditions \citep[see][for example]{Wood2010}. Non-neutral stability conditions arise when the vertical virtual temperature profile deviates from adiabatic conditions, and this affect the atmospheric boundary-layer depth as well as velocity, temperature and turbulence properties in urban areas.

Wind tunnel simulations that include the effects of approaching flow stratification are rare. Only a few studies can be listed, since only few facilities worldwide are capable of simulating non-neutral flows and the development of the correct experimental methodology is very time-consuming. The work by \citet{Uehara2000} is one of the most cited experimental works on thermal stratification, also thanks to the wide range of stability levels tested. It deals with an array of aligned cubes, where stratification ranged from a bulk Richardson number based on the building height ($H$) $Ri_H$ of $-0.21$ to $0.79$. A two-component laser-Doppler anemometer (LDA) coupled with a cold wire (CW) were employed to measure velocity and temperature. Measurements were all performed by scanning the central vertical cross-section downstream a cube with only one profile extending up to an height of $z/H = 7$ (boundary layer top, $\delta$). The stable boundary layer (SBL) was found to reduce the velocity inside and above the canopy, until the point where real stagnation regions formed at the bottom (for $Ri_H$ larger than 0.4), hence altering the entire cavity eddy. A convective boundary layer (CBL) was found to have opposite effects, strengthening the downward velocities and the reverse flow in the canopy. This resulted in vertical mixing which reduced the temperature difference, hence weakening buoyancy. Shear stresses and turbulence in the canopy were found largely sensitive to stratification, with such effects extending also to the internal boundary layer (IBL) forming over the canopy (estimated to reach a height of $2.5H$).

Another very interesting experimental work is the one conducted by \citet{Kanda2016} with a staggered array of cubes (with height either equal to the base edge or double this length). Both a SBL and a CBL (bulk Richardson number based on the boundary layer top $Ri_\delta$ = 0.21 and $-0.27$, respectively) were tested against a neutral boundary layer (NBL). Velocity and heat fluxes were acquired with a three-component LDA coupled with a cold wire for only a single vertical profile in the canopy. The analysis of the velocity and temperature statistics revealed how the flow was stratified also inside the canopy, determining a sensible reduction of turbulence in case of SBL and an increment for CBL. A constant-flux layer (CFL) region was identified above the canopy, whose extension was increased in case of a CBL. Concentration measurements were also performed, revealing a clear stratification effect even close to the source.

Quite a few studies investigated arrays of buildings with buoyancy forces employing numerical simulations with the large-eddy simulation (LES) technique. \citet{Inagaki2012} simulated a full-height daytime CBL over a square array of aligned cubes with imposed ground and roof heat flux. They showed that the turbulent organised structures above the canopy are correlated to the strong upward motion that occurs within the cavity of the arrays. A similar geometry and methodology was considered by \citet{Park2013}. In the bottom-heating case, they observed plume-shaped structures appearing together with the streamwise-elongated ones, which determined an increment in the magnitude of vertical turbulent momentum flux, partly due to ejections.

Staggered arrays of cubes were studied by \citet{Xie2013} and \citet{Boppana2014}. In the former, mean velocity and turbulent statistics were set at the inlet with a turbulence generator while the surfaces were adiabatic. $Ri_H$ ranged from $-$0.2 to 0.2. Results showed that the velocity fluctuation field was found to differ more from neutral conditions in the case of CBL, while in case of SBL the block size dominated the turbulent flow. Differently, \citet{Boppana2014} achieved non-neutrally-stratified conditions by means of setting a constant heat flux, either positive or negative, on the bottom surface. They found the turbulence intensity to be significantly affected by ground heating and cooling. The turbulent integral length scales from the two-point spatial correlations were observed to be reduced in both streamwise and vertical directions by stable stratification when compared to the neutral case, while in case of ground heating only the vertical integral length scale was found to be increased.

Recent numerical simulations focussed on pollutant dispersion in urban arrays \citep{Tomas2016,Shen2017,Jiang2018}, however their results remain largely unvalidated due to the lack of experimental data.
\citet{Nazarian2016} and \citet{Nazarian2018} considered an array of aligned cubes and rectangular buildings, respectively. Contrary to most of the previous literature, they studied a case with realistic non-uniform heating of all the surfaces in the model. They stressed the importance of considering a three-dimensional heating for studies of thermal comfort. At the same time, the concentration field was mainly affected by the overall heating of the surfaces and a detailed three-dimensional heating was found superfluous in this regard.

The literature analysis highlighted a lack of experimental works on the topic of urban stratification, with only two works dealing with stratified approaching flows and uniform array of buildings. In this paper we develop a slightly different strategy for the generation of the non-neutral boundary layers, as described in detail by \citet{Marucci2018}. The ratio $H/\delta$ is also more realistic compared to \citet{Uehara2000} thanks to the use of spires at the inlet \citep[a similar approach was used by][]{Kanda2016}. Another element of novelty in the present study is the use of a more realistic urban geometry, while still using a regular array of buildings. Compared to previous studies on regular arrays of cubes, the dimensions of the building blocks here are $H\times 2H\times H$, introducing both geometrical asymmetry and a minimum stretch to develop proper street canyon flows, as shown by the work of \cite{Castro2017} and \cite{Fuka2018} for neutral stratification. This paper deals mainly with the modifications in the flow, turbulence and temperature fields introduced by a non-neutral stratification on an urban array. Dispersion measurements were also carried out, but the results will be reported elsewhere \citep{Marucci2020}

\section{Methodology}
\subsection{Wind tunnel and model}
\begin{figure}
	\centering
	\includegraphics[width=\linewidth]{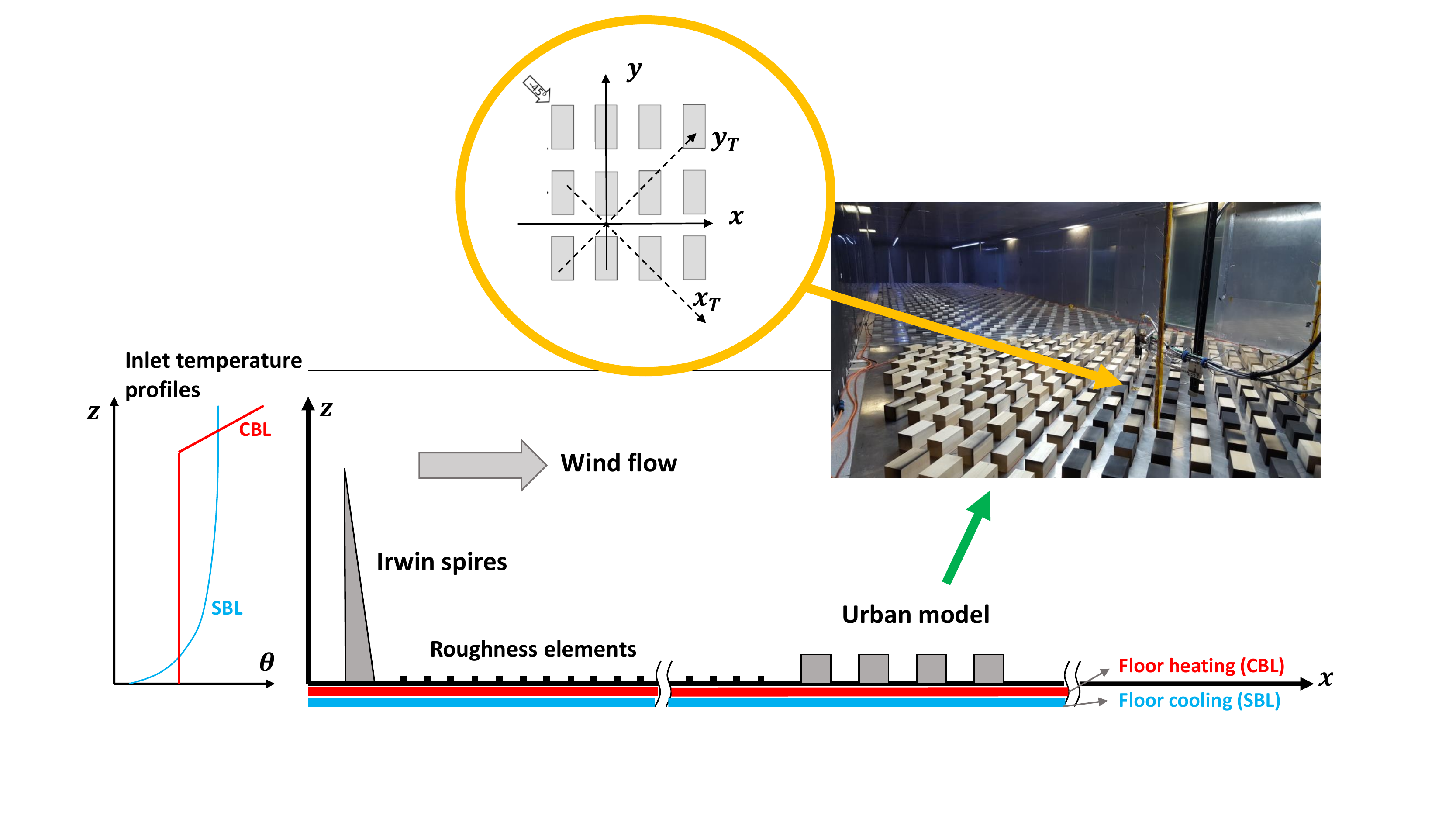}
	\caption{Schematic of the experimental set up with details about the wind tunnel coordinate system $(x_T,y_T)$ and the model coordinate system $(x,y)$.}
	\label{fig:schematic}
\end{figure}

The experiments were carried out in the EnFlo meteorological wind tunnel at the University of Surrey. The open-return facility is characterised by a working section 20~m long, 3.5~m wide and 1.5~m high. Two different sets of Irwin's spires \cite{Irwin1981} were employed to artificially thicken the boundary layer in the stable and convective cases. The first were 986~mm high, 121~mm wide at the base and 4~mm at the tip, laterally spaced 500~mm. The second were 1260~mm high and 170~mm wide at the base, laterally spaced 630~mm. Rectangular-shaped sharp-edged roughness elements were also placed on the floor in a staggered arrangement, 240~mm apart laterally and 240~mm spaced streamwise, equal for the two configurations. This was to guarantee the development of a rough approaching flow for the model. When stratified boundary layers were simulated, a vertical temperature profile was imposed at the inlet section by means of a series of fifteen 100~mm-high horizontal heaters. In the SBL cases a negative surface heat flux was generated with floor-cooling panels by means of recirculating water. In the CBL cases the floor was heated by means of electrical mats added on top of additional insulating panels. Their maximum power was 2.0~kW/m$^2$, with dimensions 1295$\times$333$\times$5~mm. Panel temperatures were controlled in a closed-loop system by means of thermistors. The details about the development and optimisation of the experimental techniques were reported by \citet{Marucci2018}, while a schematic of the experimental set up is shown in Fig.~\ref{fig:schematic}.

An ultrasonic anemometer was employed in order to provide a reference velocity ($U_{REF}$) as input for the closed-loop wind tunnel speed control system; it was placed 5~m from the inlet section, 1~m on the side of the centreline, 1~m from the floor.

Two different coordinate systems were employed (see Fig.~\ref{fig:schematic}). In the first, called ``wind tunnel coordinate system'', the $x$-axis was in the streamwise direction (along the wind tunnel centreline), measured 14~m from the working-section inlet; the $y$-axis was in the lateral direction; the $z$-axis represented the vertical, starting from the floor. The second, called ``model coordinate system'' was integral with the model, so that when the latter was rotated with respect to the wind direction, the model system was rotated by the same angle along the $z$-axis.

The urban model was constituted by a regular array of more than 350 rectangular ($H\times2H\times H$) blocks with equal height and spacing ($H = 70$~mm). Such configuration represents a significant departure from the more studied cube array, introducing a geometrical asymmetry which makes it more typical of street canyons in real urban regions \citep{Castro2017}, but at the same time remaining an organised and regular geometry. All the experiments reported here were performed using a wind direction of 45 degrees.

\subsection{Measurements}
The measurement setup was identical to the one used by \citet{Marucci2019}, to which we refer the reader for more details. Temperatures and two components of velocity were measured simultaneously using, respectively, a fast-response cold-wire probe (CW) and a laser Doppler anemometer (LDA). The LDA target acquisition frequency was set to 100~Hz, while temperatures were sampled at 1000~Hz. Given the irregular nature of the LDA measurements and the different frequencies, a resampling and synchronisation of the three signals was necessary for computing heat fluxes \citep{Marucci2019}. The measurement rig also included a fast flame ionisation detector (FFID) to measure tracer concentrations, even though concentration and mass flux results are not reported here \citep[see, e.g.][]{Marucci2019,Marucci2020}.

The averaging time for each measurement was set to 2.5 minutes, in line with previous work both in neutral \citep{Castro2017} and non-neutral \citep{Marucci2018,Marucci2019} conditions. The standard error for first and second order statistics was evaluated at each measurement point. In stable conditions standard errors for mean streamwise velocities ($\overline{U}$) were generally below 2\%, while spanwise ($\overline{V}$) and vertical ($\overline{W}$) components were around 10\% (with very small measured values, generally). Mean temperatures ($\overline{\Theta}$) had standard errors close to 0 (0.1\% on average). Variance values were generally quite low (around 5\%) for velocity components ($\overline{u'^2}$, $\overline{v'^2}$ and $\overline{w'^2}$) and temperatures ($\overline{\theta'^2}$). Neutral and, especially, convective conditions had the effect of increasing the standard errors, suggesting that future CBL experiments might be conducted with larger averaging times. CBL values tended to be between 50\% and 100\% higher than SBL standard errors. Standard errors for covariance values ($\overline{u'v'}$, $\overline{u'w'}$, $\overline{u'\theta'}$, $\overline{v'\theta'}$ and $\overline{w'\theta'}$) were generally higher and between 10 and 25\%, with little sensitivity to different stratification conditions. In the previous discussion and throughout the paper, capital letters and overbars represent a time averaged value, while small letters and the prime symbol identify fluctuating components.

\subsection{Scaling and estimation of surface properties}\label{sec:scaling}

Following the Monin-Obukhov theory for the surface layer \citep{Monin1954}, two vertical logarithmic profiles can be derived for both velocity and temperature in the general diabatic case:

\begin{equation} \label{u_mo}
\overline{U}(z) = \frac{u_\ast}{k}\left[\ln \left(\frac{z-d}{z_{0}}\right)-\psi_m\left(\zeta\right)\right]
\end{equation}

\begin{equation} \label{t_mo}
\overline{\Theta}(z) - \Theta_0 = \frac{\theta_\ast}{k}\left[\ln \left(\frac{z-d}{z_{0h}}\right)-\psi_h\left(\zeta\right)\right]
\end{equation}

\noindent where $u_\ast$ is the friction velocity, $k$ is the Von Karman constant (0.40), $d$ is the displacement height, $z_0$ is the roughness length, $\Theta_0$ is a reference temperature close to the ground, $\theta_\ast = -\left(\overline{w'\theta'}\right)_0/u_\ast$ is a scaling temperature and $z_{0h}$ is the thermal roughness length. $\psi_m$ and $\psi_h$ account for different stability conditions and are functions of the scaling ratio $\zeta = z/L$ (with $L$ being the Monin-Obukhov length) defined as:

\begin{equation} \label{zeta}
\zeta = \frac{z}{L} = -\frac{\left(g/\Theta_0\right)\theta_\ast}{u^2_\ast/kz}
\end{equation}

Following the approach by \citet{Hogstrom1988}, summarised by \citet{Marucci2018}, equations \ref{u_mo} and \ref{t_mo} can be simplified to

\begin{equation} \label{u_mo_SBL}
\overline{U}(z) = \frac{u_\ast}{k}\left[\ln \left(\frac{z-d}{z_0}\right)+8\frac{z-d-z_0}{L}\right]
\end{equation}

\begin{equation} \label{t_mo_SBL}
\overline{\Theta}(z) - \Theta_0 = \frac{\theta_\ast}{k}\left[\ln \left(\frac{z-d}{z_{0h}}\right)+16\frac{z-d-z_{0h}}{L}\right]
\end{equation}

\noindent for the SBL, while in the CBL

\begin{multline} \label{u_mo_CBL}
\overline{U}(z) = \frac{u_\ast}{k}\left[\ln \left(\frac{z-d}{z_0}\right)-\ln\left(\frac{(1+\alpha)^2}{(1+\alpha_0)^2}\frac{(1+\alpha^2)}{(1+\alpha_0^2)}\right)\right] \\
+2\frac{u_\ast}{k}\left(\tan^{-1}(\alpha)-\tan^{-1}(\alpha_0)\right)
\end{multline}

\noindent with $\alpha = \left[1-16(z-d)/L\right]^{1/4}$ and $\alpha_0 = \left(1-16z_0/L\right)^{1/4}$;

\begin{equation} \label{t_mo_CBL}
\overline{\Theta}(z) - \Theta_0 = \frac{\theta_\ast}{k}\left[\ln\left(\frac{\beta-1}{\beta+1}\right)-\ln\left(\frac{\beta_0-1}{\beta_0+1}\right)\right]
\end{equation}

\noindent with $\beta = \left[1-16(z-d)/L\right]^{1/2}$ and $\beta_0 = \left(1-16z_{0h}/L\right)^{1/2}$.\\

Two bulk Richardson numbers are considered in this paper: $Ri_\delta$ and $Ri_H$, calculated as

\begin{equation} \label{bulkRi}
Ri_\delta = \frac{g\left(\Theta_\delta - \Theta_0\right)\delta}{\Theta_0 U_\delta^2}, \ \
Ri_H = \frac{g\left(\Theta_H - \Theta_0\right)H}{\Theta_0 U_H^2}
\end{equation}

\noindent where $\Theta_\delta$, $U_\delta$, $\Theta_H$ and $U_H$ have been obtained by averaging the available samples.

The above scaling is considered valid for the surface layer, while in the mixed-layer of a CBL, following \citet{Kaimal1994}, the relevant length scale is the boundary-layer depth ($\delta$) and velocity and temperature scales are, respectively

\begin{equation}
w_\ast = \left[\frac{g}{\Theta_0}\left(\overline{w'\theta'}\right)_0 \delta\right]^{1/3}, \ \
\tilde{\theta} = \frac{\left(\overline{w'\theta'}\right)_0}{w_\ast}
\end{equation}

\begin{figure}
	\centering
	\includegraphics[width=\linewidth]{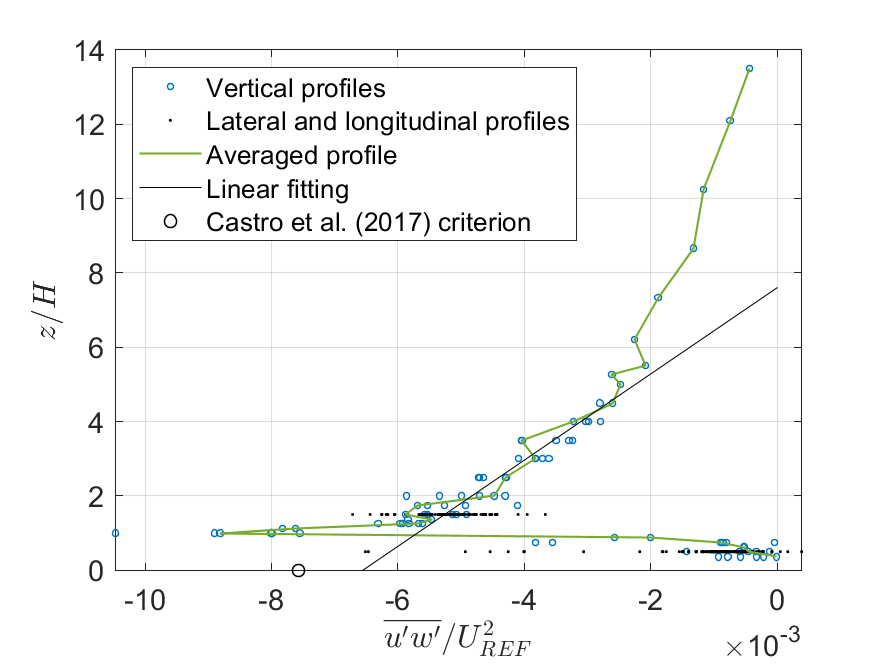}
	\caption{Reynolds shear stress at surface extrapolation for NBL case with $5\times1260$~mm spires and wind direction $45^\circ$ over urban model.}
	\label{fig:uwsurfaceNBL}
\end{figure}

Surface quantities in stratified UBLs were estimated following the methodology described by \cite{Marucci2018}. It assumed that the spatially averaged profiles for the Reynolds shear stress and vertical heat flux were approximately linear in both the roughness sub-layer (RSL) and the region immediately above. With this hypothesis (experimentally observed in that case) the value at the surface for both quantities may be obtained by means of a linear fitting of the data above the RSL alone, without the necessity of a strict spatial averaging of the profiles, being the quantities in that region independent of the local effect of the roughness. The application of this method is a necessity in the present case, as the data acquired do not have a resolution high enough for a proper spatial averaging.
A different approach was employed by \cite{Castro2017} for the determination of the friction velocity in a NBL over the same urban array of buildings tested here. They calculated the friction velocity by increasing the shear stress obtained just above the canopy by a factor of 1.3, following \cite{Cheng2002a}, obtaining $u_\ast/U_{REF}$ equal to 0.089.
Fig.~\ref{fig:uwsurfaceNBL} shows the results from the application of both methods for the NBL case and a $45^\circ$ wind direction,  employing $5\times1260$~mm spires. Four vertical profiles are available, even though only one for the entire BL depth. A linear fitting was attempted in the region between $1.5H$ and $4H$ leading to the value at the surface correspondent to a friction velocity $u_\ast/U_{REF}$ of 0.081. Such interval extends from part of the RSL (ending at about $z/H = 2$, according to \citealp{Castro2017}) up to the region above. To be noted that a constant flux layer is not discernible, as expected, likely due to the non-zero pressure gradient (see \citealp{Marucci2018}). As support of this argumentation, \cite{Kanda2016} was able to obtain a constant flux layer over an array of cubic buildings, but employing a wind tunnel with adjustable ceiling height and shape to generate a zero-pressure gradient boundary layer. Nevertheless, a reasonably good linear trend, similar to what was found in \cite{Marucci2018} for the approaching flow, is appreciable, suggesting the applicability of the same method also in this case. On the other hand, multiplying the averaged value of the shear stress at $z/H = 1.25$ for 1.3 leads to a friction velocity of 0.087, slightly higher than the value obtained with the linear fitting, but closer to the 0.089 reported by \cite{Castro2017}.
Following these considerations and because of the quite good agreement obtained with both approaches, the linear fitting method was deemed more practical in this case, also considering that no multiplicative factors are available for the vertical heat flux.

As far as the displacement height is concerned, \cite{Castro2017} suggested the values of $0.59H$ for wind directions 45$^\circ$. Such value was obtained from LES and DNS (direct numerical simulation) assuming $d$ as the height at which the surface drag appears to act \citep{Jackson1981}. They claimed that calculating $d$ together with $z_0$ by means of fitting the log-law with the streamwise wind profile and fixing the von-Karman constant $k$ would lead to larger values of $d$ and unrealistically smaller values of $z_0$. However, the latter approach was applied in the present study, as it is the most widely used approach and led to consistent results, with values of $z_0$ comparable with  \cite{Castro2017}. Finally, the same approach was considered for the thermal roughness length $z_{0h}$ calculation, in which the temperature measured at a height of 10~mm at the centre of the model was used as reference $\Theta_0$ for all the data above the array. Two cases were considered, one in which the displacement height for the temperature profile $d_h$ was maintained equal to $d$, and the other in which it was a free fitting parameter.

\section{Effects of a stable boundary layer}

\subsection{Simulated SBL characteristics}
In order to generate the two weaker stable approaching flows, the same wind tunnel settings as in \cite{Marucci2018} were employed, characterised by a non-uniform inlet temperature profile and reference velocity $U_{REF} = 1.25$~m/s. The stratification strength was adjusted by modifying $\Delta\Theta_{MAX}$, defined as the difference between the target floor temperature and the free-stream temperaure set at the inlet. The values indicated in Tab.~\ref{table:SBLpar} and labelled as ``lower roughness'' were obtained without the urban array and averaging three profiles at different longitudinal positions in the centreline within the region where the model would be located. The third, and strongest, simulated SBL was obtained by reducing the velocity to 1.15~m/s and increasing $\Delta\Theta_{MAX}$. The combined effect of velocity reduction and $\Delta\Theta_{MAX}$ increment was considered the best compromise between the need of maintaining a sufficiently high Reynolds number while preventing an overheating of the LDA probe. Unfortunately, no measurements were performed with this stratification level before placing the model (i.e., only ``higher roughness'' values are present). It should be noted that $Ri_\delta^{app}$ in the table is the nominal (or desired) bulk Richardson number of the approaching flow, which sometimes differs slightly from the one actually measured (also reported in the table). Moreover, the reference temperature $\Theta_0$ for the lower roughness data is sampled at about 2~mm from the ground, 1~m upstream of the location where the model is placed.

\begin{table}
	\caption[SBL cases parameters, ``Lower roughness'' case acquired with roughness elements only, ``Higher roughness'' case with the urban array at wind direction 45$^\circ$.]{SBL cases parameters, ``Lower roughness'' case acquired with roughness elements only, ``Higher roughness'' case with the urban array at wind direction 45$^\circ$.}
	\centering
	\label{table:SBLpar}
	\scalebox{0.75}{
		\begin{tabular}{l c c c | c c c c}
			\toprule
			& \multicolumn{3}{c|}{\textbf{Lower roughness}} & \multicolumn{4}{c}{\textbf{Higher roughness}} \\
			\midrule
			$\mathrm{Ri_\delta^{app}}$ & 0 & 0.14 & 0.21 & 0 & 0.14 & 0.21 & 0.29 \\
			$\mathrm{\Delta\Theta_{\text{MAX}}\ (^\circ C)}$ & 0 & 10.8 & 16 & 0 & 10.8 & 16 & 17.8 \\
			$\mathrm{U_{\text{REF}}\ (m/s)}$ & 1.25 & 1.25 & 1.25 & 1.25 & 1.25 & 1.25 & 1.15\\
			\midrule
			$\mathrm{u_\ast/U_{\text{REF}}}$ & 0.065 & 0.053 & 0.047 & 0.078 & 0.063 & 0.061 & 0.059 \\
			$\mathrm{z_0\ (mm)}$ &2.2&  2.3 & 2.3  & 3.45 & 2.5 & 2.6 & 2.9\\
			$\mathrm{d\ (mm)}$ & 0 & 0 & 0 & 52.5 & 53.5 & 54.5 & 55.0 \\
			$\mathrm{\delta\ (mm)}$ & 850 & 850 & 850 & 850 & 850 & 850 & 850 \\
			$\mathrm{\Theta_0\ (^\circ C)}$ & - & 16.3 & 17.0 & - & 17.4 & 17.8 & 18 \\
			$\mathrm{\Delta\Theta \left[=\Theta_{\delta}-\Theta_0\right]}$ & - & 9.3 & 13.5 & - & 8.2 & 12.8 & 14.3 \\
			$\mathrm{\theta_\ast\ (^\circ C)}$ &-& 0.24 & 0.34  & - & 0.221 & 0.315 & 0.355\\
			$\mathrm{z_{0h}\ (mm) \left[d_h=d\right]}$ & - & 0.007 & 0.012 & - & 0.006 & 0.004 & 0.006\\
			$\mathrm{d_h\ (mm) \left[Fitted\right]}$ & - & - & - & - & 51.4 & 47.3 & 37.4\\
			$\mathrm{z_{0h}\ (mm) \left[d_h\ fitted\right]}$ & - & - & - & - & 0.006 & 0.004 & 0.010\\
			\midrule
			$\mathrm{\delta/L}$ & 0 & 0.63 & 1.13 & 0 & 0.40 & 0.62 & 0.88\\
			$\mathrm{Ri_\delta}$ & 0 & 0.15 & 0.21 & 0 & 0.12 & 0.19 & 0.24\\
			$\mathrm{Ri_H}$ & 0 & 0.035 & 0.054 & 0 & 0.10 & 0.19 & 0.28\\
			$\mathrm{Re_\ast}$ & 10.2 & 8.7 & 7.7 & 22.7 & 11.2 & 13.3 & 11.8\\
			$\mathrm{Re_\delta\ \left(x10^3\right)}$ & 75.90 & 77.45 & 77.98 & 67.09 & 78.57 & 79.95 & 74.30\\
			\bottomrule
	\end{tabular}}
\end{table}

A comparison between the lower roughness values in Tab.~\ref{table:SBLpar} and the ones over the urban array allows to understand how the scaling parameters vary due to the increased roughness. Firstly, the data show how the friction velocity is gradually reduced by applying a stable stratification both in the lower and higher roughness cases. However, while the reduction with $Ri_\delta^{app}=0.14$ is about 19\% in both cases, when $Ri_\delta^{app}=0.21$ the friction velocity is only 22\% lower than the NBL above the canopy, compared to a 28\% reduction experienced when the urban array is not in place. Moreover, a further increase in stability produce just an additional 2\% reduction. This suggests that while stratification affects the flow above the canopy, an increment in roughness tends to reduce the sensitivity of the friction velocity to the imposed stratification variations.

While in the lower roughness case no significant stratification effects on the aerodynamic roughness length $z_0$ were observed \citep{Marucci2018}, with the urban array there is a reduction between 16 and 27\%. At the same time, the displacement height $d$ shows a slight gradual increase with increasing stratification (up to 5\% larger) to a value around 0.8$H$ \citep[similar to][for cubes and other geometries]{Jackson1981}. On the other hand, \cite{Uehara2000} did not find any significant effect after applying stable stratification (incidentally, not even in the CBL case), despite having comparable values for $z_0$ and $d$.

The boundary layer depth $\delta$ does not seem to be affected by the presence of the model, as can be observed from the Reynolds shear stress plot in Fig.~\ref{fig:SBL021over}. Both the methods used to evaluate the thermal roughness parameters (see section \ref{sec:scaling}) bring to similar results, with differences in displacement height $d_h$ of less than 15\% and a $z_{0h}$ of the same order of magnitude, comparable with the one measured in the lower roughness case.

The $\Delta\Theta$ differences, combined with a slight increment of the velocity due to the blockage above the model (the latter being less than 3\%), brings to a reduction of the measured $Ri_\delta$ between 10 and 17\%. Differently, the modification in the Monin-Obukhov length $L$ is clearly larger, with an increment up to 80\% due to the increased roughness, which causes a reduction of the surface stability. Table \ref{table:SBLpar} also reports the value of $Ri_H$, where the averaged values of velocity and temperature measured at roof level are considered. It is interesting to note that above the array $Ri_\delta \approx Ri_H$, meaning that the velocity reduction at canopy top compared to the BL top compensates for the lower temperature difference and reference length. The same is not true for the other case, where the lower roughness causes a larger increment of velocity closer to the ground. Finally, two Reynolds numbers are also reported: $Re_\ast=u_\ast z_0/\nu$ and $Re_\delta=U_\delta \delta/\nu$, where $\nu$ is the air kinematic viscosity at $\Theta_0$.

\subsection{SBL flow above the canopy}
\begin{figure}
	\centering
	\includegraphics[width=\linewidth]{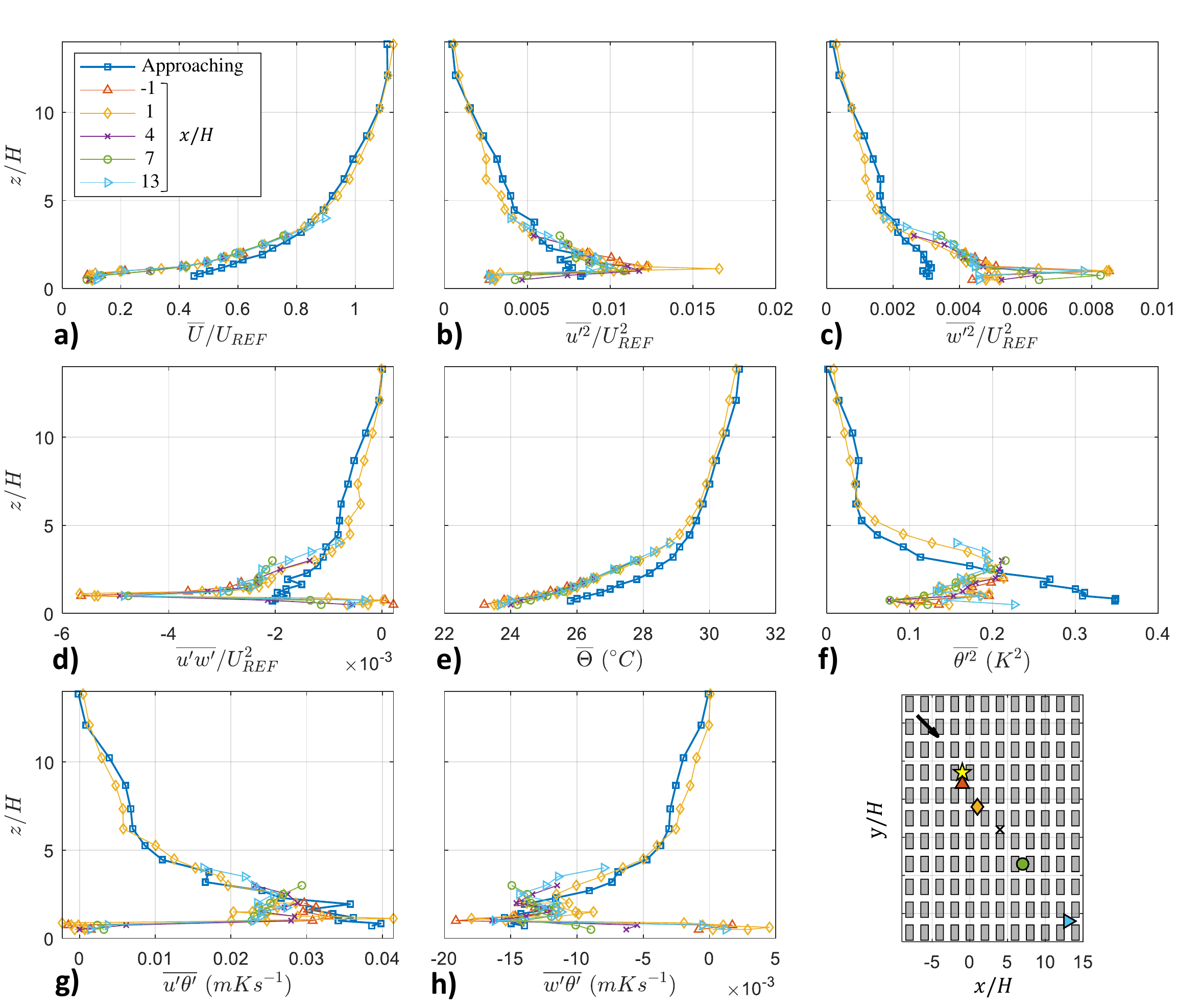}
	\caption{Vertical profiles of first and second order statistics for velocity and temperature at $Ri_\delta^{app}=0.21$. The approach profile is sampled at $x_T/H = -35$, about 1.5~m upstream of the model.}
	\label{fig:SBL021over}
\end{figure}

The vertical profiles presented in Fig.~\ref{fig:SBL021over} were located in the second half of the array where, according to the work done in the same model by \citet{Castro2017}, the NBL would be fully developed. Regarding the SBL development, clear modifications take place above the array compared to the approaching flow (measured at $x_T/H = -35$, about 1.5~m upstream the model), while in the second half of the array the values in the first 25-30\% of the BL do not differ too much at the various locations.
The mean streamwise velocity appears slower than the approaching flow for the bottom quarter of the BL due to the increased drag, and faster above. Consequently, the power law coefficient $\alpha$ increases, varying from 0.40 to 0.60. The latter was computed by fitting $\overline{U}/U_R=\left(z/z_R\right)^\alpha$ with the mean velocity profile (in which $U_R$ is the velocity sampled at $z_R=0.4\delta$). Curiously, the value of $\alpha$ above the array in neutral stratification is again 0.40, suggesting that for this parameter the applied stable stratification has an effect similar to the increment in roughness under neutral conditions.

The Reynolds shear stresses above 3$H$ appear only slightly modified by the presence of the building array, suggesting a similar depth of the internal boundary layer developing above the canopy. Below, on the other hand, a dramatic increase is experienced, peaking at roof level. Interestingly, above 3$H$ the turbulence over the array appears slightly reduced compared to the approaching flow, despite the opposite effect of the higher roughness.
As far as the mean temperature is concerned, the measured values over the array are lower than in the approaching flow up to 5$H$, unchanged above. The temperature gradient extends clearly into the canopy with a larger gradient than above. The temperature variance profile, on the other hand, extends its similarity region down to 2.5$H$, below which the fluctuations are heavily damped. The streamwise and vertical heat fluxes appear less sensitive to the increase of roughness compared to the other quantities.

\begin{figure}
	\centering
	\includegraphics[width=0.7\linewidth]{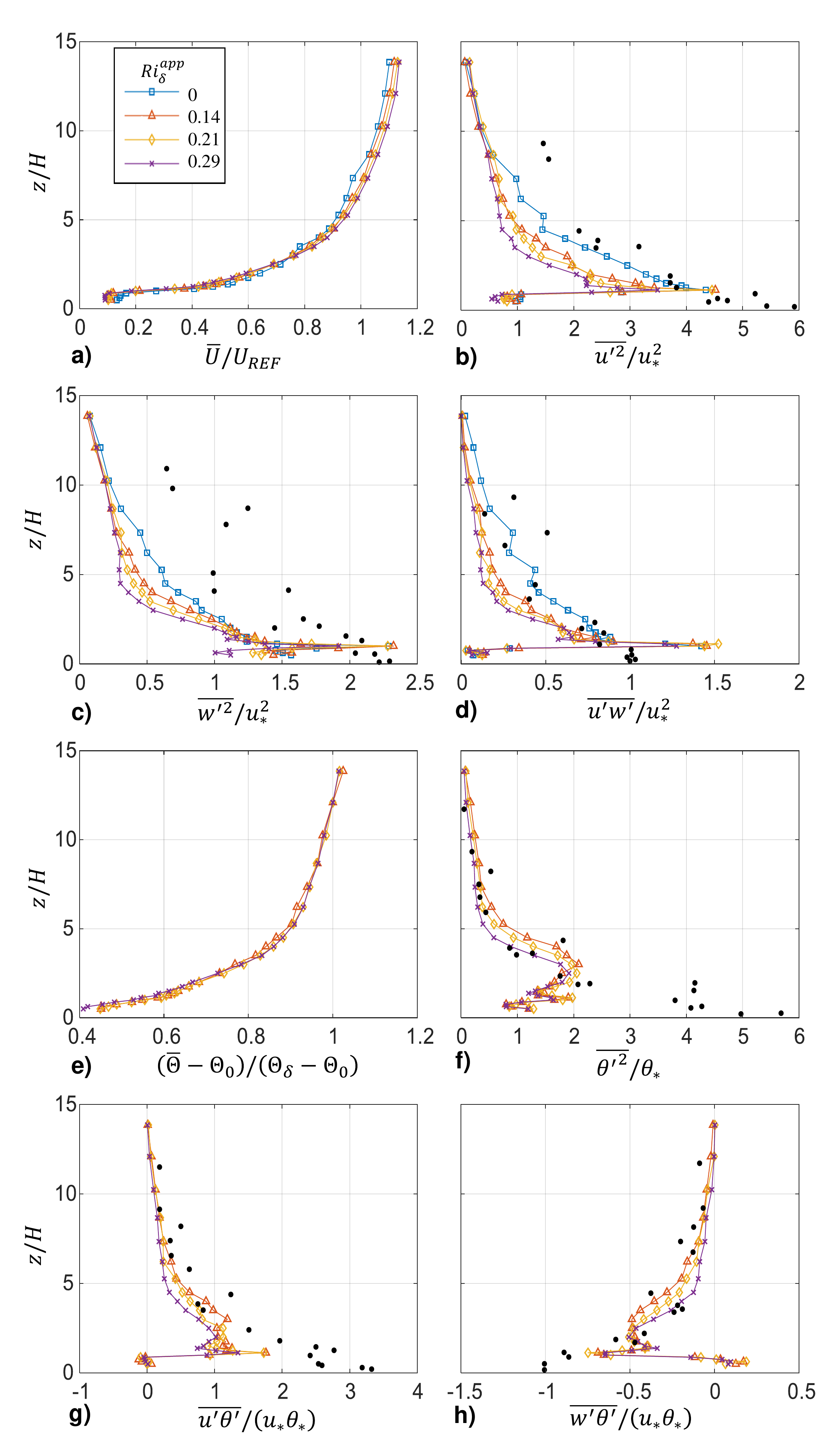}
	\caption{Vertical profiles of first and second order statistics for velocity and temperature varying the level of stability at $x/H = 1$, $y/H = -6$. Black dots are field data from \cite{Caughey1979}.}
	\label{fig:SBLArraynonDim}
\end{figure}

Fig.~\ref{fig:SBLArraynonDim} presents a comparison between the three different stable stratification cases and the reference neutral at one location. The Reynolds stresses are non-dimensionalised by the friction velocity, while in the quantity involving the temperature fluctuation the friction temperature $\theta_\ast$ is used. In the mean velocity graph it is possible to appreciate how the slope varies with stability layer strength: the values are gradually larger above 2.5$H$ and lower below, causing an increment of the $\alpha$ coefficient.
For the Reynolds stresses, the turbulence reduction due to stratification causes the profile to change shape, becoming more curved, and thus deviating from the neutral case \citep[as already observed by][for roughness conditions similar to the approaching flow]{Marucci2018}.
The mean temperature profile appears almost unchanged in shape when varying stratification (by means of changing the $\Delta\Theta$). The temperature variance graphs show a peak at around 2.5$H$, slightly reducing in height with increasing stratification and followed below by an almost linear reduction. The heat flux graphs have a similar behaviour, but with a small region of constant flux above the canopy and a more marked peak at roof level, even though these quantities depend heavily on location in the roughness sub-layer.

As reference the field data from \cite{Caughey1979} are also plotted (Fig.~\ref{fig:SBLArraynonDim}). For the Reynolds stresses, their trend is quite linear with height, causing the SBL data to deviate as stratification increases (due to the already mentioned increment in the curvature of the profile). Not surprisingly the NBL, looking more linear, seems to fit better the field data. The largest difference seems to be for the vertical velocity fluctuations, likely caused by a very high value of the friction velocity due to the large roughness imposed. The agreement is clearly better for the thermal quantities, which appear to follow the trend, at least above 2.5$H$. It is interesting to note that for a cubic array of blocks under stratified and neutral conditions \cite{Uehara2000} found a height of the internal boundary layer of 2.5$H$, as in the present case.

\begin{figure}
	\centering
	\includegraphics[width=\linewidth]{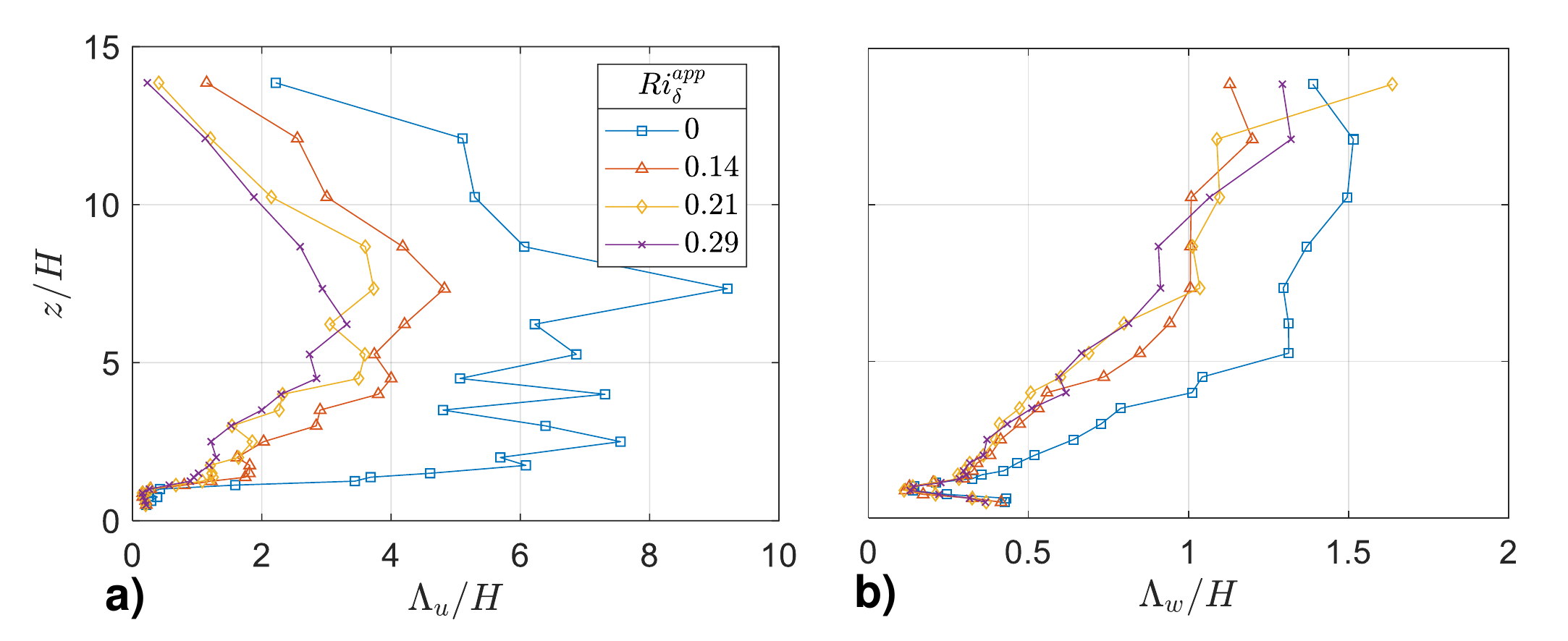}
	\caption{Vertical profiles of integral length scales for SBL cases at $x/H = 1$, $y/H = -6$}
	\label{fig:LengthScaleSBL}
\end{figure}

Integral length scales for the streamwise and vertical velocity are also reported (Fig.~\ref{fig:LengthScaleSBL}), computed from the numerical integration of the autocorrelation coefficient, assuming the Taylor's hypothesis of ``frozen turbulence''. The streamwise velocity length scale for the neutral case has a large scatter, with the profile increasing up to about 7$H$ (0.5$\delta$) and then decreasing. The amount of the length scale at the peak is also around 7$H$, hence closer to what indicated by \cite{Robins1979} and \cite{Shirakata2002} ($\approx 0.3\delta$) than the values found by \cite{Kanda2016} ($\approx 1.1\delta$). The vertical velocity length scale, differently, increases almost monotonically with the height above the canopy, but remaining confined to lower values (up to 1.5$H$). In the stable cases all the length scales preserve the same trend as in the neutral, but with progressively smaller values.

\subsection{Flow inside the canopy and channelling effects}
\begin{figure}
	\centering
	\includegraphics[width=0.7\linewidth]{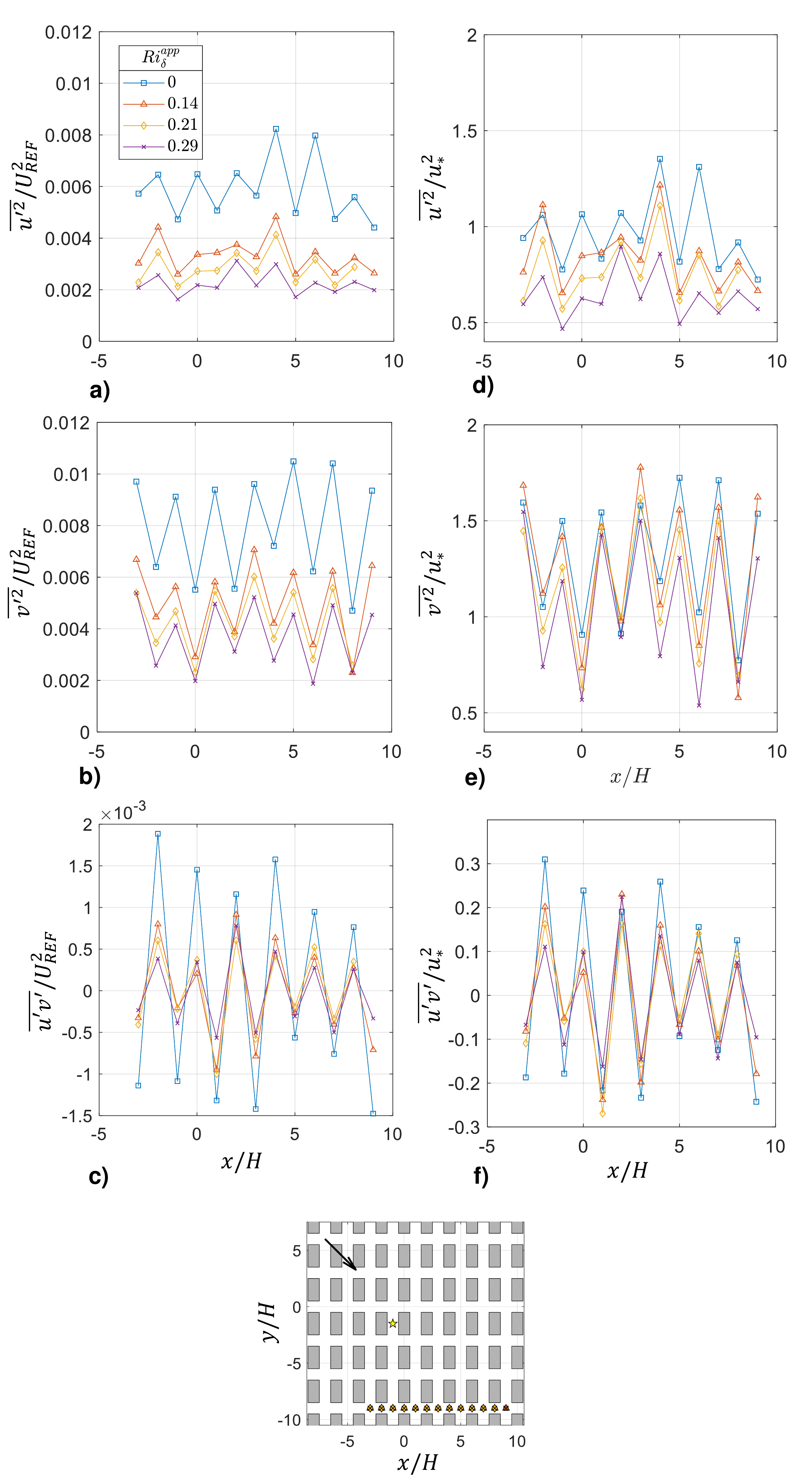}
	\caption{Reynolds stresses inside the canopy  ($z/H = 0.5$, $y/H = -9$) varying the stable stratification. Quantities are non-dimensionalised by both the reference (a-c) and friction (d-f) velocity.}
	\label{fig:SBLinside}
\end{figure}

Fig.~\ref{fig:SBLinside}a-c shows the Reynolds stresses inside the canopy on the streets facing the short-edge of the buildings. Despite the high level of turbulence mechanically produced by the building blocks, a clear and gradual reduction due to stratification is perfectly appreciable with values up to four times lower than the neutral case. In  Fig.~\ref{fig:SBLinside}d-f the same quantities are presented, but non-dimensionalised by the friction velocity. They do not seem to scale perfectly according to this parameter, with the SBL values systematically smaller than the NBL ones (as it was indeed above the canopy, too).

\begin{figure}
	\centering
	\includegraphics[width=\linewidth]{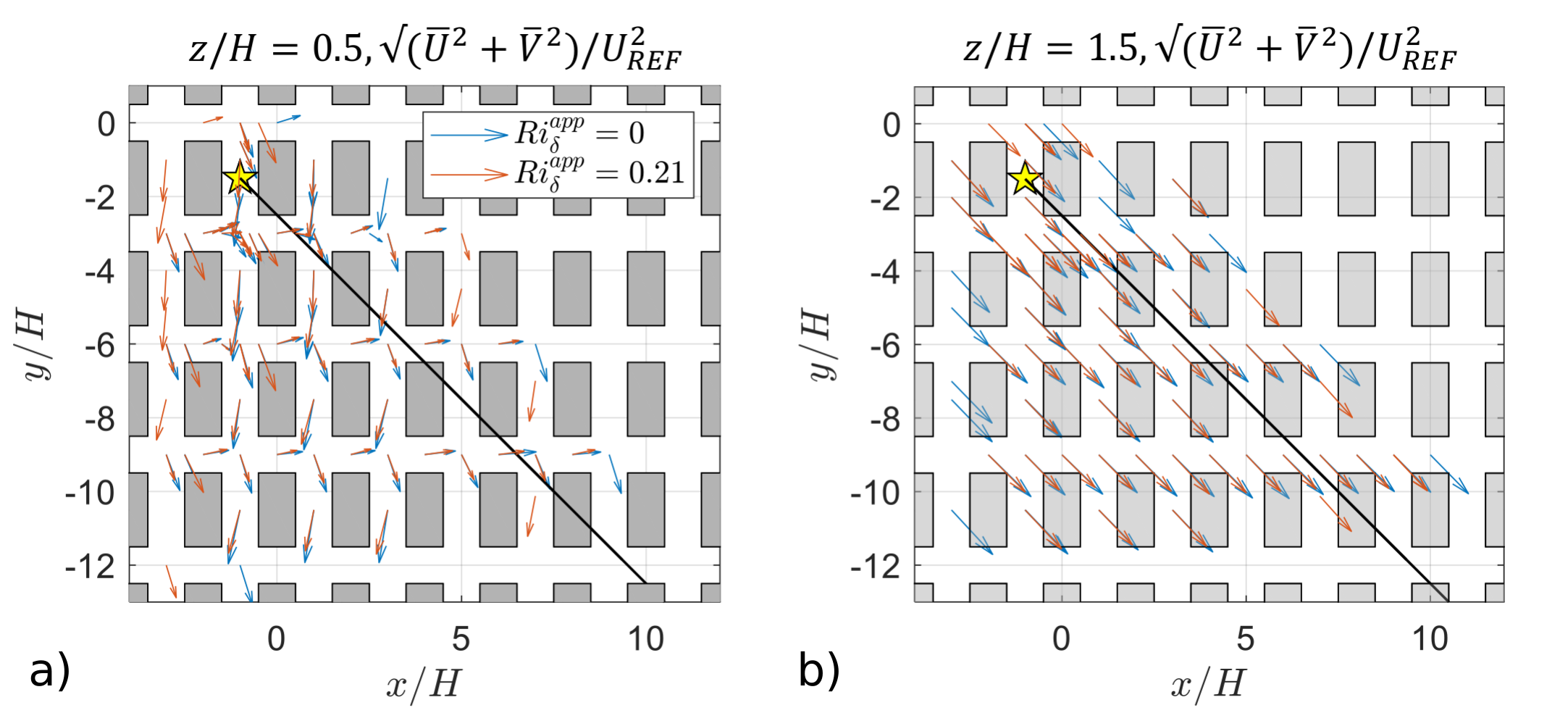}
	\caption{Planar view of mean horizontal velocity vectors inside ($z/H = 0.5$, left) and above ($z/H = 1.5$, right) the canopy for SBL and NBL. The star symbol represents the source location for the dispersion study reported by \citet{Marucci2020}.}
	\label{fig:SBLVelInside}
\end{figure}

In terms of flow channelling, the chosen urban array model was found to produce a street canyon type flow \citep{Castro2017}, even though the ratio between long and short edge of the building was only 2. This means that the velocity in the street facing the long edge of the buildings is expected to approximately align with the road centreline, hence deviating from the mean flow direction above the canopy. Such trend is clearly visible in Fig.~\ref{fig:SBLVelInside}, where vectors of horizontal velocity are plotted at $z/H = 0.5$. The channelling appears well developed already in the NBL, while the addition of stable stratification does not seem to increase this trend. Nevertheless, the main effect of the stratification on the mean velocity inside the canopy is a general reduction of the magnitude, as already noted by other authors, see e.g. \cite{Uehara2000}, \cite{Li2016}, \cite{Kanda2016}. The former explained this behaviour by the fact that cavity eddies developing inside the street canyon would be weakened by SBL (the opposite for CBL) with a reduced downward flow which, for larger Richardson numbers (not investigated here), would result in nearly zero velocity inside the canyons. Half building height above the canopy, instead, the flow is already perfectly aligned with the free-stream direction.

\section{Effects of a convective boundary layer}

\subsection{Simulated CBL characteristics}
Tab.~\ref{table:CBLpar} reports the main parameters for the CBL simulations. A uniform temperature profile was set at the inlet, capped by a linear inversion of roughly 10$^\circ$~C/m starting from 1~m upwards (as detailed in \citealp{Marucci2018}). Comparing the values reported in Tab.~\ref{table:SBLpar} and \ref{table:CBLpar}, there are some small differences in the reference NBL parameters, mainly due to the different sets of spires employed.

\begin{table}
	\caption[CBL cases parameters, wind direction 45$^\circ$.]{CBL cases parameters, wind direction 45$^\circ$.}
	\centering
	\label{table:CBLpar}
	\scalebox{0.75}{
		\begin{tabular}{l c c c | c c c}
			\toprule
			& \multicolumn{3}{c|}{\textbf{Lower roughness}} & \multicolumn{3}{c}{\textbf{Higher roughness}} \\
			\midrule
			$\mathrm{Ri_\delta^{app}}$ & 0 & $-$0.5 & $-$1.5 & 0 & $-$0.5 & $-$1.5 \\
			$\mathrm{\Delta\Theta_{\text{MAX}}\ (^\circ C)}$ & 0 & $-$24.2 & $-$39.2 & 0 & $-$24.2 & $-$39.2 \\
			$\mathrm{U_{\text{REF}}\ (m/s)}$ & 1.25 & 1.25 & 1.0 & 1.25 & 1.25 & 1.0\\
			\midrule
			$\mathrm{u_\ast/U_{\text{REF}}}$ & 0.067 & 0.088 & 0.101 & 0.081 & 0.105 & 0.118\\
			$\mathrm{z_0\ (mm)}$ &2.0&  2.2 & 2.1 & 4.0 & 6.3 & 6.2\\
			$\mathrm{d\ (mm)}$ & 0 & 0 & 0 & 50.8 & 23.5 & 21.5\\
			$\mathrm{\delta\ (mm)}$ & 1000 & 1200 & 1350 & 1000 & 1200 & 1350\\
			$\mathrm{\Theta_0\ (^\circ C)}$ & - & 44.4 & 59.0 & - & 39.0 & 50.0\\
			$\mathrm{\Delta\Theta \left[=\Theta_{\delta}-\Theta_0\right]}$ & - & $-$21.4 & $-$34.0 & - & $-$15.8 & $-$24.6\\
			$\mathrm{\theta_\ast\ (^\circ C)}$ &-& $-$0.81 & $-$1.39  & - & $-$0.60 & $-$0.92\\
			$\mathrm{w_\ast/U_{REF}}$ &-& 0.119 & 0.177 & - & 0.115 & 0.158\\
			$\mathrm{z_{0h}\ (mm) \left[d_h=d\right]}$ & - & 0.0045 & 0.0037 & - & 0.0067 & 0.0037\\
			$\mathrm{d_h\ (mm) \left[Fitted\right]}$ & - & - & - & - & 52.3 & 44.5\\
			$\mathrm{z_{0h}\ (mm) \left[d_h\ fitted\right]}$ & - & - & - & - & 0.0050 & 0.0030\\
			\midrule
			$\mathrm{\delta/L}$ & 0 & $-$0.99 & $-$2.18 & 0 & $-$0.51 & $-$1.09\\
			$\mathrm{u_\ast/w_\ast}$ & - & 0.74 & 0.57 & - & 0.92 & 0.75\\
			$\mathrm{Ri_\delta}$ & 0 & $-$0.50 & $-$1.44 & 0 & $-$0.35 & $-$0.91\\
			$\mathrm{Ri_H}$ & 0 & $-$0.057 & $-$0.078 & 0 & $-$0.15 & $-$0.19\\
			$\mathrm{Re_\ast}$ & 12.4 & 17.7 & 11.3 & 26.8 & 49.5 & 40.8\\
			$\mathrm{Re_\delta\ \left(x10^3\right)}$ & 88.5 & 89.8 & 73.2 & 87.8 & 92.7 & 74.6 \\
			\bottomrule
	\end{tabular}}
\end{table}

The effect of a CBL on the friction velocity is analogous to an increment in roughness. In fact, the value of $u_\ast$ over the array with a NBL (0.081) is similar to the one in the lower roughness with $Ri_\delta^{app} = -0.5$ (0.088) and the same can be said comparing $Ri_\delta^{app} = -0.5$ over the array (0.105) with the case with the strongest convective conditions but lower roughness (0.101). Moreover, as it was found for the SBL, the effect of increment in friction velocity consequent to the application of the stratification is less marked over the array (with an increase of roughly 45\% respect to the NBL case) than in the lower roughness case (where the increment was more consistent, 50\%).

Regarding the effect of the unstable stratification on $z_0$ and $d$, it can be noted that the former experiences an increase of approximately 55\% from the NBL, regardless the strength of the convective conditions, while the latter appears reduced by roughly one half. The boundary layer depth was assumed to remain the same above the array as in the case without. In fact, the value of $\delta$ over the array  cannot be too different from the one in the lower roughness case (estimated in \citealp{Marucci2018}) due to height constrains of the wind tunnel ceiling. On the other hand, a more accurate estimation was not possible due to limitations in the measuring range.

The vertical heat flux over the array appears reduced, also as consequence of the wooden buildings not being heated. The values of thermal roughness length are all very close to each other (same order of magnitude), and so are the displacement heights $d_h$ determined from the temperature profile. The increment in friction velocity combined with the reduction of vertical heat flux causes an increase in the values of Monin-Obukhov length over the array which are doubled compared to the approaching flow. A similar behaviour is also found for the $u_\ast/w_\ast$ ratio and $Ri_\delta$ indicating a clear reduction of the strength of the convective conditions. The values of $Ri_H$ in the unstable cases are lower than the respective $Ri_\delta$, also as a consequence of the fact that the difference between $\delta$ and $H$ is much larger for the CBLs compared to the SBLs, in which the two Richardson numbers were found comparable.

\subsection{CBL flow above the canopy}
\begin{figure}
	\centering
	\includegraphics[width=\linewidth]{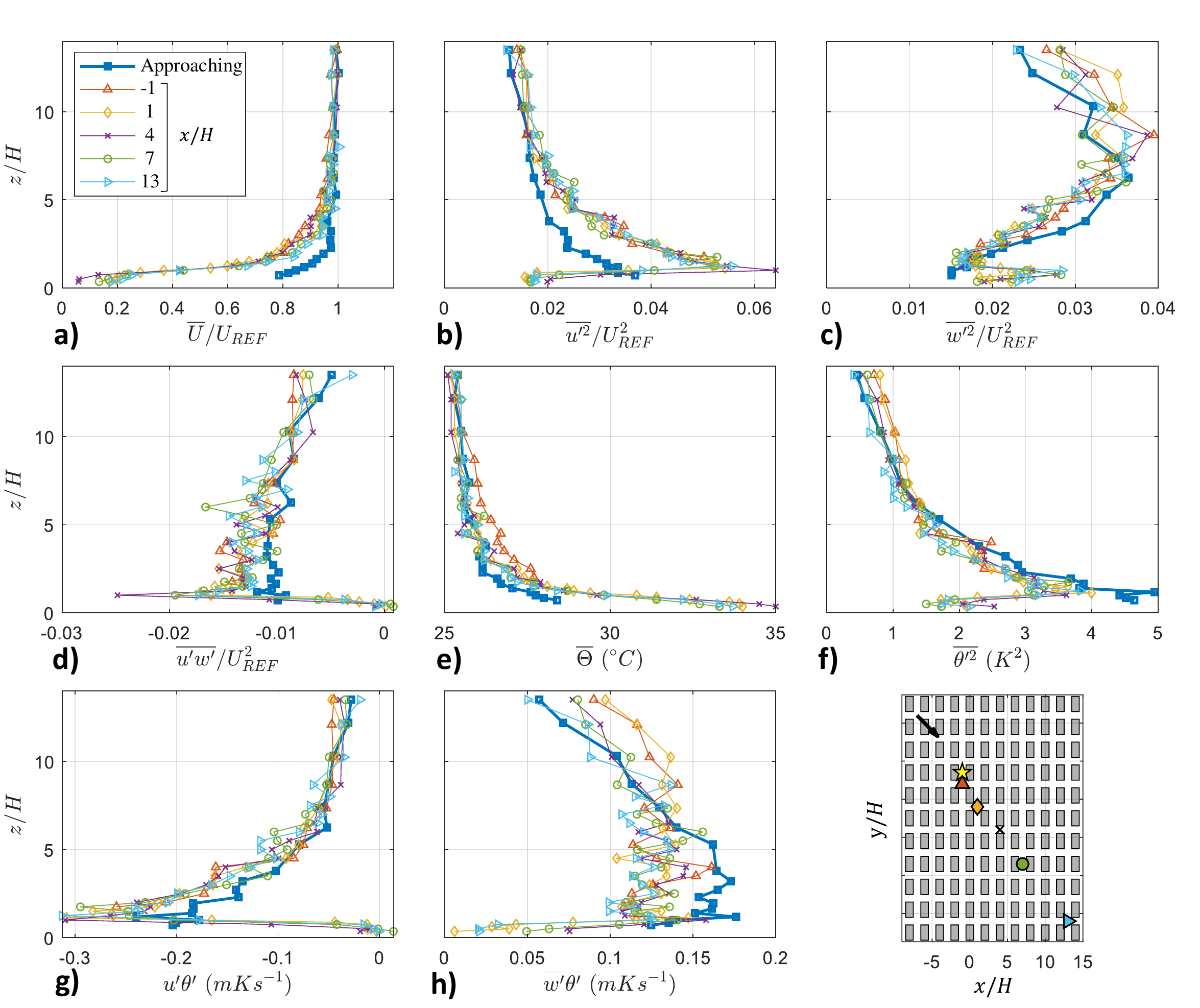}
	\caption[Vertical profiles of first and second order statistics for velocity and temperature for $Ri_\delta^{app}=-1.5$.]{Vertical profiles of first and second order statistics for velocity and temperature for $Ri_\delta^{app}=-1.5$. The approaching data has been sampled without the urban array and obtained as average of two profiles at $x_T/H = 1.4$ and 22.4, $y_T/H = 0$ in wind tunnel coordinates.}
	\label{fig:CBLm1_5over}
\end{figure}

In Fig.~\ref{fig:CBLm1_5over} the vertical profiles for the lower and higher roughness cases are compared for the $Ri_\delta^{app}=-1.5$ experiments. The average of two profiles acquired at $x_T/H = 1.4$ and 22.4 without the model is considered. The reason for the difference with the SBL cases, where a single profile at $x_T/H=-35$ was used (see Fig.~\ref{fig:SBL021over}), is that at that point the CBL is still not sufficiently developed, and so it would not represent a fair comparison. As a general observation, the various profiles above the array show a good degree of similarity, meaning that the flow reached an equilibrium with the roughness underneath and is not evolving longitudinally too much.
On the other hand, the differences between the approaching flow and the vertical profiles above the urban model are significant. The flow slows down as an effect of the increased roughness for heights up to 4$H$, while above that no differences are found on the velocities. Since for the conservation of mass the flow rate has to remain constant, there should be a further acceleration in the region above 1~m, but no measurements were performed at such heights. As an effect of the reduction in velocity, the mean temperature is higher in the canopy and immediately above, while the temperature fluctuations are unchanged.

Streamwise velocity variance is larger above the canopy up to about 7$H$ while no significant differences are experienced by the vertical component in the same region, despite the increase in roughness, meaning that, as expected, the mixed layer does not scale with the friction velocity. The Reynolds shear stress is slightly increased over the array (as also indicated by the friction velocity). The opposite happens for the vertical heat flux which, despite the larger temperature gradient, appears reduced. As already mentioned, a possible reason is the fact that the wooden buildings are colder than the heated panels. On the other hand, the streamwise temperature flux is slightly increased over the array

\begin{figure}
	\centering
	\includegraphics[width=0.7\linewidth]{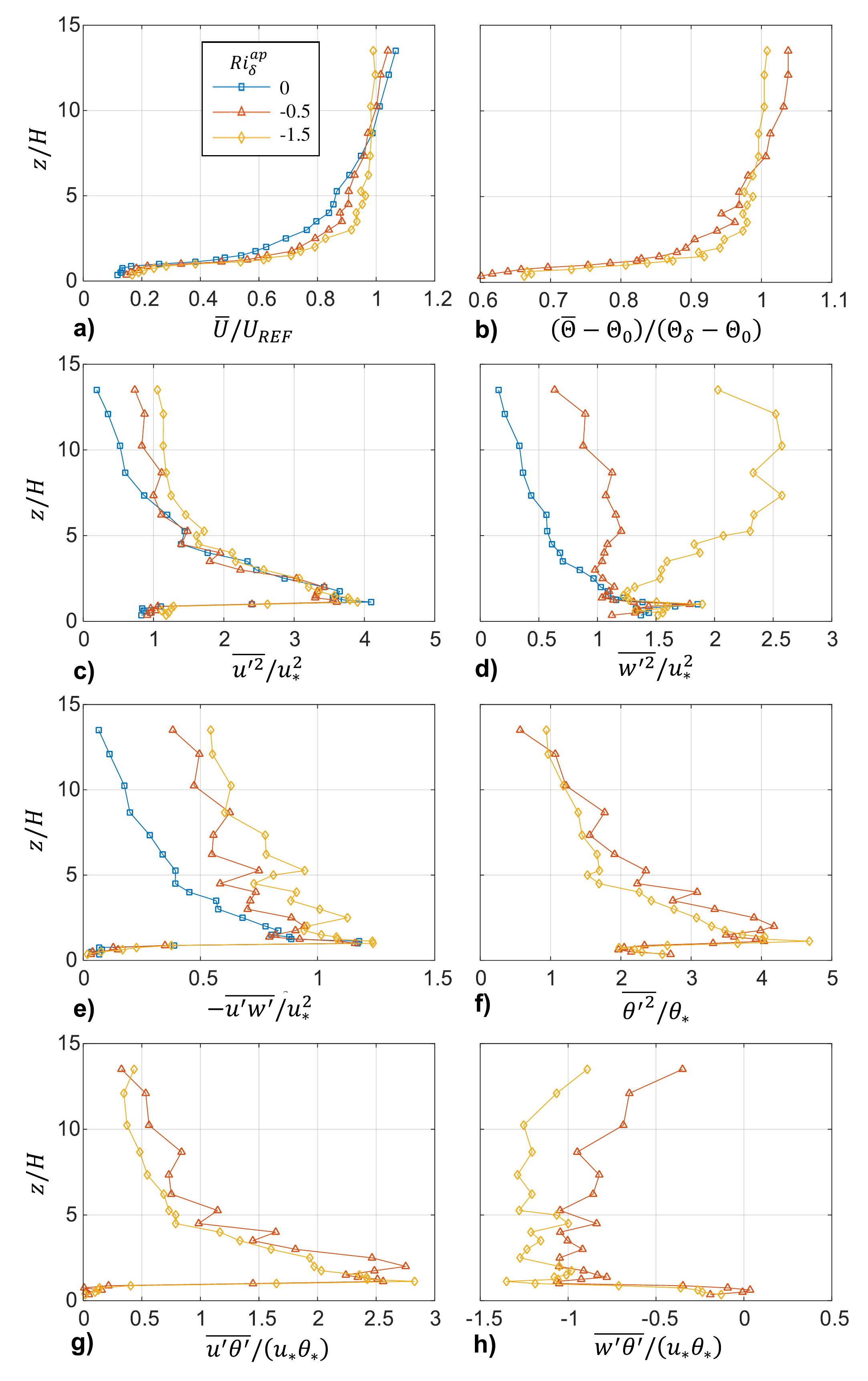}
	\caption{Vertical profiles of first and second order statistics for velocity and temperature varying the strength of the convective conditions at $x/H = 1$, $y/H = -6$.}
	\label{fig:CBLArraynonDim}
\end{figure}

Fig.~\ref{fig:CBLArraynonDim} compares results from different stratification levels at one location. The axes are non-dimensionalised by the friction velocity and temperature. As the unstable stratification increases a mixed layer develops above the roughness sub-layer, with almost constant velocity above $z/H = 3$ in the most unstable case. For the $Ri_\delta^{app}=-0.5$ case a velocity profile that is of intermediate shape between the NBL and the $Ri_\delta^{app}=-1.5$ cases is observed. A similar consideration can be made for the mean temperature profile. The streamwise velocity variance appears to scale appropriately with the friction velocity up to 5$H$, while above this the unstable stratification causes an increase in fluctuations. The threshold appears to be lower for the vertical velocity variance, where the vertical profile starts to be different immediately above the canopy. Observation of the Reynolds shear stress reveals that a region of strictly constant flux above the canopy never develops, as also observed by \cite{Cheng2002a}. On the other hand, vertical heat fluxes present a constant region up to about 8-10$H$. Finally, streamwise heat fluxes and temperature fluctuations scales reasonably well with friction temperature.

\begin{figure}
	\centering
	\includegraphics[width=\linewidth]{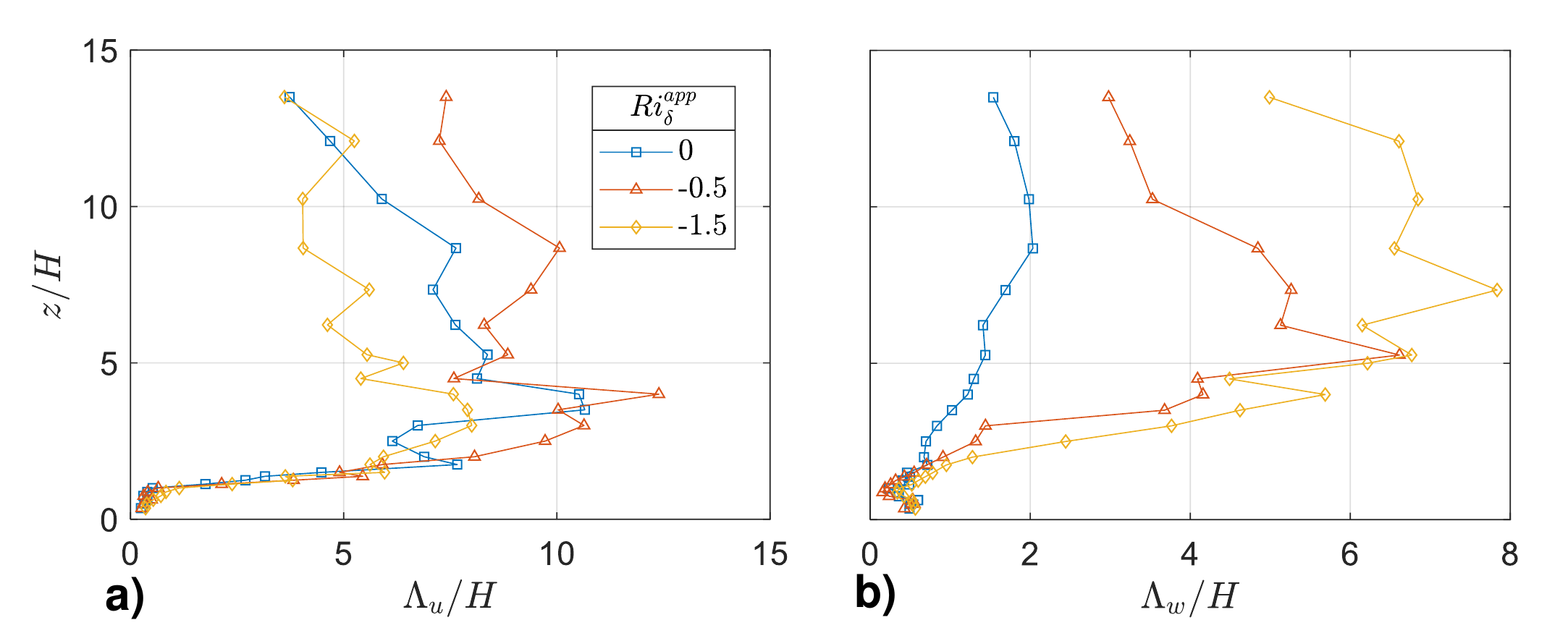}
	\caption{Vertical profiles of integral length scales for CBL cases at $x/H = 1$, $y/H = -6$.}
	\label{fig:LengthScaleCBL}
\end{figure}

The integral length scales for the streamwise and vertical velocity are reported in Fig.~\ref{fig:LengthScaleCBL}. For the streamwise length scale the $Ri_{\delta}^{app} = -0.5$ case shows quite a similar trend compared to the neutral case, with only slightly larger values above $z/H = 5$. Differently, the case $Ri_{\delta}^{app} = -1.5$ shows slightly smaller values compared to the neutral one. This apparently opposite trend is very likely just due to the large scatter. It should be noted that also \cite{Boppana2014} found only a very small reduction in the streamwise velocity length scale after the application of CBL. On the other hand, in the vertical component of the velocity the length scales increase with the strength of the convective conditions, as expected, due to the larger vertical structure and boundary layer depth. Peaks of $\Lambda_w$ are, respectively, 0.14$\delta$, 0.3$\delta$ and 0.36$\delta$ for the three cases considered. The differences in the vertical velocity length scale are much larger than in the streamwise one. On this aspect, by comparing experiments with and without spires, \cite{Kanda2016} found that $\Lambda_u$ is very dependent on the employed turbulence generator, more than on the applied stratification.

\begin{figure}
	\centering
	\includegraphics[width=\linewidth]{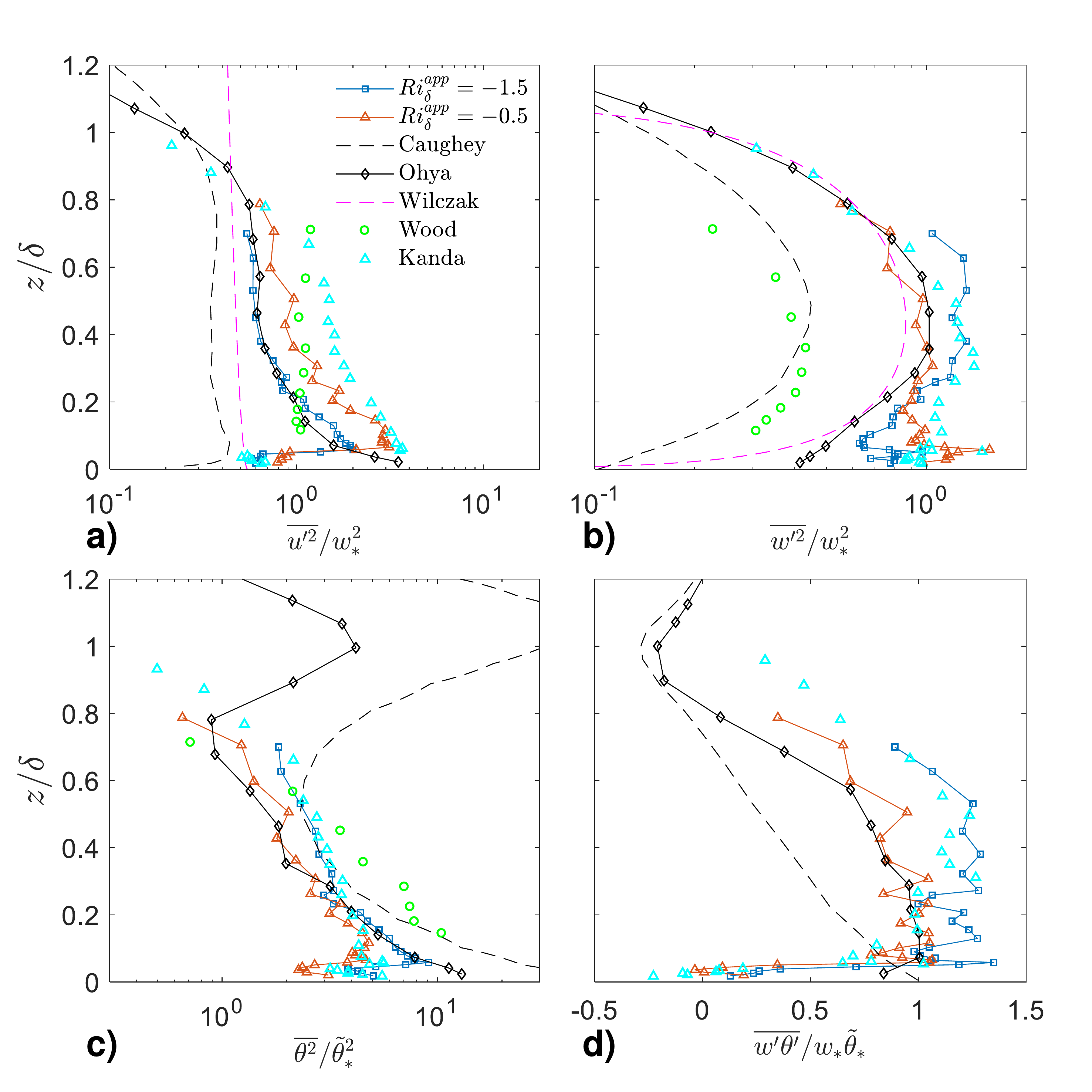}
	\caption{Profiles of non-dimensional Reynolds stresses, temperature variance and vertical kinematic heat flux at $x/H = 1$, $y/H = -6$. Data is compared with \cite{Caughey1979a}, \cite{Ohya2004} (case E2), \cite{Wilczak1986}, \cite{Wood2010} and \cite{Kanda2016}.}
	\label{fig:CBLvertwtilde}
\end{figure}

Finally, velocity and temperature variances as well as vertical heat fluxes are shown in Fig.~\ref{fig:CBLvertwtilde} normalised by the mixed layer scaling velocity and temperature. They are compared with data from literature. Profiles above the canopy do not differ much from the ones without model, as already noted in Fig.~\ref{fig:CBLm1_5over}, hence similar comments to the ones provided in \cite{Marucci2018} apply also here. In addition, experimental profiles by \cite{Kanda2016} are also considered (they refer to a CBL case characterised by $Ri_\delta = -0.27$ and $\delta/L = -0.61$). The trend they show for all the turbulent quantities is remarkably similar to the one found here, in particular for the vertical heat flux. \cite{Kanda2016} commented on the difference between the heat flux profile measure in the laboratory and the linear trend from field measurements, attributing the discrepancy to the usage of fences and spires, which introduce larger energetic eddies. Differently, in the SBL cases, the larger eddies are suppressed.

\subsection{CBL flow inside the canopy}
\begin{figure}
	\centering
	\includegraphics[width=0.7\linewidth]{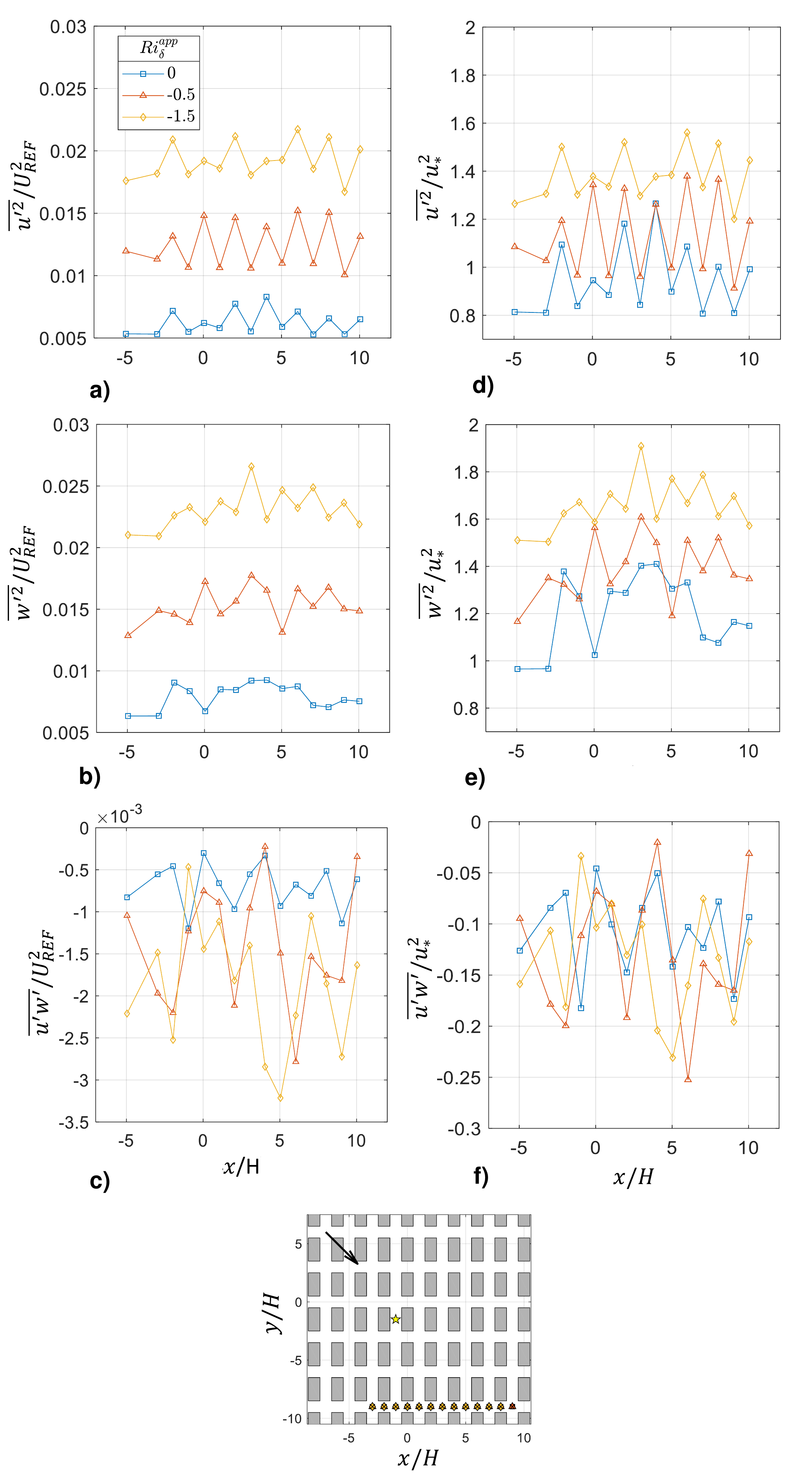}
	\caption{Reynolds stresses inside the canopy  ($z/H = 0.5$, $y/H = -9$) varying the unstable stratification. Quantities are non-dimensionalised by both the reference (a-c) and friction (d-f) velocity.}
	\label{fig:CBLinside}
\end{figure}

Fig.~\ref{fig:CBLinside} shows the Reynolds stresses inside the canopy. The application of an incoming unstable boundary layer lead to a clear increase in the magnitude of the streamwise and vertical velocity variance. For $\overline{u'w'}$ the scatter between the points at different locations but same position respect to local buildings in the CBL cases suggests a high level of unsteadiness, for which a longer measuring time would be preferable. Fig.~\ref{fig:CBLinside}d-f presents the same graphs non-dimensionalised by the friction velocity.

\section{Conclusion}
\label{Conclusion}

An experimental campaign aiming to investigate the effects of atmospheric stratification on flow over an aligned array of rectangular blocks with a wind angle of 45 degrees was performed in the EnFlo wind tunnel. A series of three stable and two convective boundary layers was employed, together with reference neutral cases, with Richardson number of the approaching flow ranging from $-$1.5 to 0.29. Velocities and temperatures were sampled using, respectively, a bi-component LDA and a cold-wire placed close to each other, allowing the point measurement of mean and fluctuating quantities, as well as Reynolds shear stresses and heat fluxes. Measurements were performed inside and above the canopy, as well as measurements of the undisturbed approaching flow, to evaluate the effect of the presence of the model. Dispersion measurements were also performed simultaneously and results are reported elsewhere \citep{Marucci2020}.

As far as stable stratification is concerned, results on the flow above and inside the canopy show a clear reduction of the Reynolds stresses (which reflects in a reduction of the friction velocity), despite the high level of roughness. The latter, however, caused an increment of the Monin-Obukhov length up to 80\% compared to the approaching flow. The aerodynamic roughness length and displacement height seem affected by stratification, with a reduction up to 27\% for the former and an increment up to 5\% for the latter. On the other hand, the wind direction of the flow inside and immediately above the canopy are not influenced, even though the mean values appear reduced. A clear reduction of the turbulence within the canopy was observed. Comparisons between the approaching flow and boundary layer over the canopy suggest a height of the internal boundary layer of about 2.5$H$, in agreement with what \cite{Uehara2000} found for an array of cubes.

In the convective stratification cases, the friction velocity appears increased by both the effect of roughness and unstable stratification, even though the sum of the two contributions considered singularly is larger than the increment resulting by their combined effect. As it was for the stable case, the increased roughness causes a reduction in the surface stratification, reflected in an increase of the Monin-Obukhov length, which is double over the array compared to the approaching flow. The effect on the aerodynamic roughness length and displacement height are specular to the SBL case, an increase up to 55\% of the former and a reduction of the same amount for the latter. The observation of the mean velocity profile suggests a height of the internal layer between 3 and 4$H$, invariant along $x$ in the measurement region.

This work helps shedding more light on the effects of stratification in the urban environment and encourages further work on the topic. The experimental database produced during the project is unique and of high quality. It can assist in developing, improving and validating numerical models, as well as developing parametrisations for simpler models.

\section*{Acknowledgments}
This work was funded by the EPSRC (grant EP/P000029/1) and by the Department of Mechanical Engineering Sciences (University of Surrey). The authors confirm that all wind tunnel data are fully available without restriction from \href{https://doi.org/10.6084/m9.figshare.8320007}{https://doi.org/10.6084/m9.figshare.8320007}.





\bibliography{StratEnFlo_array_flow_v2}

\begin{thebibliography}{30}
\expandafter\ifx\csname natexlab\endcsname\relax\def\natexlab#1{#1}\fi
\providecommand{\url}[1]{\texttt{#1}}
\providecommand{\href}[2]{#2}
\providecommand{\path}[1]{#1}
\providecommand{\DOIprefix}{doi:}
\providecommand{\ArXivprefix}{arXiv:}
\providecommand{\URLprefix}{URL: }
\providecommand{\Pubmedprefix}{pmid:}
\providecommand{\doi}[1]{\href{http://dx.doi.org/#1}{\path{#1}}}
\providecommand{\Pubmed}[1]{\href{pmid:#1}{\path{#1}}}
\providecommand{\bibinfo}[2]{#2}
\ifx\xfnm\relax \def\xfnm[#1]{\unskip,\space#1}\fi
\bibitem[{Boppana et~al.(2014)Boppana, Xie and Castro}]{Boppana2014}
\bibinfo{author}{Boppana, V.B.L.}, \bibinfo{author}{Xie, Z.T.},
  \bibinfo{author}{Castro, I.P.}, \bibinfo{year}{2014}.
\newblock \bibinfo{title}{{Thermal Stratification Effects on Flow Over a
  Generic Urban Canopy}}.
\newblock \bibinfo{journal}{Boundary-Layer Meteorology} \bibinfo{volume}{153},
  \bibinfo{pages}{141--162}.
\newblock \DOIprefix\doi{10.1007/s10546-014-9935-1}.
\bibitem[{Castro et~al.(2017)Castro, Xie, Fuka, Robins, Carpentieri, Hayden,
  Hertwig and Coceal}]{Castro2017}
\bibinfo{author}{Castro, I.P.}, \bibinfo{author}{Xie, Z.T.},
  \bibinfo{author}{Fuka, V.}, \bibinfo{author}{Robins, A.G.},
  \bibinfo{author}{Carpentieri, M.}, \bibinfo{author}{Hayden, P.},
  \bibinfo{author}{Hertwig, D.}, \bibinfo{author}{Coceal, O.},
  \bibinfo{year}{2017}.
\newblock \bibinfo{title}{{Measurements and Computations of Flow in an Urban
  Street System}}.
\newblock \bibinfo{journal}{Boundary-Layer Meteorology} \bibinfo{volume}{162},
  \bibinfo{pages}{207--230}.
\newblock \DOIprefix\doi{10.1007/s10546-016-0200-7}.
\bibitem[{Caughey and Palmer(1979)}]{Caughey1979a}
\bibinfo{author}{Caughey, S.J.}, \bibinfo{author}{Palmer, S.G.},
  \bibinfo{year}{1979}.
\newblock \bibinfo{title}{{Some aspects of turbulence structure through the
  depth of the convective boundary layer}}.
\newblock \bibinfo{journal}{Quarterly Journal of the Royal Meteorological
  Society} \bibinfo{volume}{105}, \bibinfo{pages}{811--827}.
\newblock \DOIprefix\doi{10.1002/qj.49710544606}.
\bibitem[{Caughey et~al.(1979)Caughey, Wyngaard and Kaimal}]{Caughey1979}
\bibinfo{author}{Caughey, S.J.}, \bibinfo{author}{Wyngaard, J.C.},
  \bibinfo{author}{Kaimal, J.C.}, \bibinfo{year}{1979}.
\newblock \bibinfo{title}{{Turbulence in the evolving stable boundary layer}}.
\newblock \bibinfo{journal}{J. Atmos. Sci.} \bibinfo{volume}{36},
  \bibinfo{pages}{1041--1052}.
\newblock \DOIprefix\doi{10.1175/1520-0469(1979)036<1041:TITESB>2.0.CO;2}.
\bibitem[{Cheng and Castro(2002)}]{Cheng2002a}
\bibinfo{author}{Cheng, H.}, \bibinfo{author}{Castro, I.},
  \bibinfo{year}{2002}.
\newblock \bibinfo{title}{{Near wall flow development after a step change in
  surface roughness}}.
\newblock \bibinfo{journal}{Boundary-layer Meteorology} \bibinfo{volume}{105},
  \bibinfo{pages}{411--432}.
\newblock \DOIprefix\doi{10.1023/A:1020355306788}.
\bibitem[{Fuka et~al.(2018)Fuka, Xie, Castro, Hayden, Carpentieri and
  Robins}]{Fuka2018}
\bibinfo{author}{Fuka, V.}, \bibinfo{author}{Xie, Z.T.},
  \bibinfo{author}{Castro, I.P.}, \bibinfo{author}{Hayden, P.},
  \bibinfo{author}{Carpentieri, M.}, \bibinfo{author}{Robins, A.G.},
  \bibinfo{year}{2018}.
\newblock \bibinfo{title}{{Scalar fluxes near a tall building in an aligned
  array of rectangular buildings}}.
\newblock \bibinfo{journal}{Boundary-Layer Meteorology} \bibinfo{volume}{167},
  \bibinfo{pages}{53--76}.
\newblock \DOIprefix\doi{10.1007/s10546-017-0308-4}.
\bibitem[{H{\"{o}}gstr{\"{o}}m(1988)}]{Hogstrom1988}
\bibinfo{author}{H{\"{o}}gstr{\"{o}}m, U.}, \bibinfo{year}{1988}.
\newblock \bibinfo{title}{{Non-Dimensional Wind and Temperature Profiles in the
  Atmospheric Surface Layer: A Re-Evaluation}}.
\newblock \bibinfo{journal}{Boundary-Layer Meteorology} \bibinfo{volume}{42},
  \bibinfo{pages}{55--78}.
\newblock \DOIprefix\doi{10.1007/978-94-009-2935-7_6}.
\bibitem[{Inagaki et~al.(2012)Inagaki, Castillo, Yamashita, Kanda and
  Takimoto}]{Inagaki2012}
\bibinfo{author}{Inagaki, A.}, \bibinfo{author}{Castillo, M.C.L.},
  \bibinfo{author}{Yamashita, Y.}, \bibinfo{author}{Kanda, M.},
  \bibinfo{author}{Takimoto, H.}, \bibinfo{year}{2012}.
\newblock \bibinfo{title}{{Large-Eddy Simulation of Coherent Flow Structures
  within a Cubical Canopy}}.
\newblock \bibinfo{journal}{Boundary-Layer Meteorology} \bibinfo{volume}{142},
  \bibinfo{pages}{207--222}.
\newblock \DOIprefix\doi{10.1007/s10546-011-9671-8}.
\bibitem[{Irwin(1981)}]{Irwin1981}
\bibinfo{author}{Irwin, H.P.A.H.}, \bibinfo{year}{1981}.
\newblock \bibinfo{title}{{The design of spires for wind simulation}}.
\newblock \bibinfo{journal}{Journal of Wind Engineering and Industrial
  Aerodynamics} \bibinfo{volume}{7}, \bibinfo{pages}{361--366}.
\newblock \DOIprefix\doi{10.1016/0167-6105(81)90058-1}.
\bibitem[{Jackson(1981)}]{Jackson1981}
\bibinfo{author}{Jackson, P.}, \bibinfo{year}{1981}.
\newblock \bibinfo{title}{{On the displacement height in the logarithmic
  velocity profile}}.
\newblock \bibinfo{journal}{J. Fluid Mech.} \bibinfo{volume}{111},
  \bibinfo{pages}{15--25}.
\bibitem[{Jiang and Yoshie(2018)}]{Jiang2018}
\bibinfo{author}{Jiang, G.}, \bibinfo{author}{Yoshie, R.},
  \bibinfo{year}{2018}.
\newblock \bibinfo{title}{{Large-eddy simulation of flow and pollutant
  dispersion in a 3D urban street model located in an unstable boundary
  layer}}.
\newblock \bibinfo{journal}{Building and Environment} \bibinfo{volume}{142},
  \bibinfo{pages}{47--57}.
\newblock \URLprefix \url{https://doi.org/10.1016/j.buildenv.2018.06.015},
  \DOIprefix\doi{10.1016/j.buildenv.2018.06.015}.
\bibitem[{Kaimal and Finnigan(1994)}]{Kaimal1994}
\bibinfo{author}{Kaimal, J.C.}, \bibinfo{author}{Finnigan, J.J.},
  \bibinfo{year}{1994}.
\newblock \bibinfo{title}{{Atmospheric boundary layer flows: their structure
  and measurement}}. volume~\bibinfo{volume}{72}.
\newblock \bibinfo{publisher}{Oxford University Press}.
\newblock \DOIprefix\doi{10.1016/0021-9169(95)90002-0}.
\bibitem[{Kanda and Yamao(2016)}]{Kanda2016}
\bibinfo{author}{Kanda, I.}, \bibinfo{author}{Yamao, Y.}, \bibinfo{year}{2016}.
\newblock \bibinfo{title}{{Passive scalar diffusion in and above urban-like
  roughness under weakly stable and unstable thermal stratification
  conditions}}.
\newblock \bibinfo{journal}{Journal of Wind Engineering and Industrial
  Aerodynamics} \bibinfo{volume}{148}, \bibinfo{pages}{18--33}.
\newblock \URLprefix \url{http://dx.doi.org/10.1016/j.jweia.2015.11.002},
  \DOIprefix\doi{10.1016/j.jweia.2015.11.002}.
\bibitem[{Li et~al.(2016)Li, Britter and Norford}]{Li2016}
\bibinfo{author}{Li, X.X.}, \bibinfo{author}{Britter, R.},
  \bibinfo{author}{Norford, L.K.}, \bibinfo{year}{2016}.
\newblock \bibinfo{title}{{Effect of stable stratification on dispersion within
  urban street canyons: A large-eddy simulation}}.
\newblock \bibinfo{journal}{Atmospheric Environment} \bibinfo{volume}{144},
  \bibinfo{pages}{47--59}.
\newblock \DOIprefix\doi{10.1016/j.atmosenv.2016.08.069}.
\bibitem[{Marucci and Carpentieri(2019)}]{Marucci2019}
\bibinfo{author}{Marucci, D.}, \bibinfo{author}{Carpentieri, M.},
  \bibinfo{year}{2019}.
\newblock \bibinfo{title}{{Effect of local and upwind stratification on flow
  and dispersion inside and above a bi-dimensional street canyon}}.
\newblock \bibinfo{journal}{Building and Environment} \bibinfo{volume}{156},
  \bibinfo{pages}{74--88}.
\newblock \URLprefix \url{http://arxiv.org/abs/1812.00512},
  \DOIprefix\doi{10.1016/j.buildenv.2019.04.013},
  \href{http://arxiv.org/abs/1812.00512}{{\tt arXiv:1812.00512}}.
\bibitem[{Marucci and Carpentieri(2020)}]{Marucci2020}
\bibinfo{author}{Marucci, D.}, \bibinfo{author}{Carpentieri, M.},
  \bibinfo{year}{2020}.
\newblock \bibinfo{title}{{Dispersion in an array of buildings in stable and
  convective atmospheric conditions}}.
\newblock \bibinfo{journal}{Atmospheric Environment} \bibinfo{volume}{222},
  \bibinfo{pages}{117100}.
\newblock \URLprefix \url{https://arxiv.org/abs/1908.06027},
  \DOIprefix\doi{10.1016/j.atmosenv.2019.117100},
  \href{http://arxiv.org/abs/1908.06027}{{\tt arXiv:1908.06027}}.
\bibitem[{Marucci et~al.(2018)Marucci, Carpentieri and Hayden}]{Marucci2018}
\bibinfo{author}{Marucci, D.}, \bibinfo{author}{Carpentieri, M.},
  \bibinfo{author}{Hayden, P.}, \bibinfo{year}{2018}.
\newblock \bibinfo{title}{{On the simulation of thick non-neutral boundary
  layers for urban studies in a wind tunnel}}.
\newblock \bibinfo{journal}{International Journal of Heat and Fluid Flow}
  \bibinfo{volume}{72}, \bibinfo{pages}{37--51}.
\newblock \DOIprefix\doi{10.1016/j.ijheatfluidflow.2018.05.012}.
\bibitem[{Monin and Obukhov(1954)}]{Monin1954}
\bibinfo{author}{Monin, A.S.}, \bibinfo{author}{Obukhov, A.M.},
  \bibinfo{year}{1954}.
\newblock \bibinfo{title}{{Basic laws of turbulent mixing in the surface layer
  of the atmosphere}}.
\newblock \bibinfo{journal}{Contrib. Geophys. Inst. Acad. Sci. USSR}
  \bibinfo{volume}{24}, \bibinfo{pages}{163--187}.
\bibitem[{Nazarian and Kleissl(2016)}]{Nazarian2016}
\bibinfo{author}{Nazarian, N.}, \bibinfo{author}{Kleissl, J.},
  \bibinfo{year}{2016}.
\newblock \bibinfo{title}{{Realistic solar heating in urban areas : Air
  exchange and street-canyon ventilation}}.
\newblock \bibinfo{journal}{Building and Environment} \bibinfo{volume}{95},
  \bibinfo{pages}{75--93}.
\newblock \URLprefix \url{http://dx.doi.org/10.1016/j.buildenv.2015.08.021},
  \DOIprefix\doi{10.1016/j.buildenv.2015.08.021}.
\bibitem[{Nazarian et~al.(2018)Nazarian, Martilli and Kleissl}]{Nazarian2018}
\bibinfo{author}{Nazarian, N.}, \bibinfo{author}{Martilli, A.},
  \bibinfo{author}{Kleissl, J.}, \bibinfo{year}{2018}.
\newblock \bibinfo{title}{{Impacts of Realistic Urban Heating , Part I :
  Spatial Variability of Mean Flow , Turbulent Exchange and Pollutant
  Dispersion}}.
\newblock \bibinfo{journal}{Boundary-Layer Meteorology} \bibinfo{volume}{166},
  \bibinfo{pages}{367--393}.
\newblock \DOIprefix\doi{10.1007/s10546-017-0311-9}.
\bibitem[{Ohya and Uchida(2004)}]{Ohya2004}
\bibinfo{author}{Ohya, Y.}, \bibinfo{author}{Uchida, T.}, \bibinfo{year}{2004}.
\newblock \bibinfo{title}{{Laboratory and numerical studies of the convective
  boundary layer capped by a strong inversion}}.
\newblock \bibinfo{journal}{Boundary-Layer Meteorology} \bibinfo{volume}{112},
  \bibinfo{pages}{223--240}.
\newblock \DOIprefix\doi{10.1023/B:BOUN.0000027913.22130.73}.
\bibitem[{Park and Baik(2013)}]{Park2013}
\bibinfo{author}{Park, S.B.}, \bibinfo{author}{Baik, J.J.},
  \bibinfo{year}{2013}.
\newblock \bibinfo{title}{{A Large-Eddy Simulation Study of Thermal Effects on
  Turbulence Coherent Structures in and above a Building Array}}.
\newblock \bibinfo{journal}{Journal of applied meteorology and climatology}
  \bibinfo{volume}{52}, \bibinfo{pages}{1348--1365}.
\newblock \DOIprefix\doi{10.1175/JAMC-D-12-0162.1}.
\bibitem[{Robins(1979)}]{Robins1979}
\bibinfo{author}{Robins, A.G.}, \bibinfo{year}{1979}.
\newblock \bibinfo{title}{{The Development and Structure of Simulated Neutrally
  Stable Atmospheric Boundary Layers}}.
\newblock \bibinfo{journal}{Journal of Industrial Aerodynamics}
  \bibinfo{volume}{4}, \bibinfo{pages}{71--100}.
\bibitem[{Shen et~al.(2017)Shen, Cui and Zhang}]{Shen2017}
\bibinfo{author}{Shen, Z.}, \bibinfo{author}{Cui, G.}, \bibinfo{author}{Zhang,
  Z.}, \bibinfo{year}{2017}.
\newblock \bibinfo{title}{{Turbulent dispersion of pollutants in urban-type
  canopies under stable stratification conditions}}.
\newblock \bibinfo{journal}{Atmospheric Environment} \bibinfo{volume}{156},
  \bibinfo{pages}{1--14}.
\newblock \URLprefix \url{http://dx.doi.org/10.1016/j.atmosenv.2017.02.017},
  \DOIprefix\doi{10.1016/j.atmosenv.2017.02.017}.
\bibitem[{Shirakata et~al.(2002)Shirakata, Nagai and Mizumoto}]{Shirakata2002}
\bibinfo{author}{Shirakata, S.}, \bibinfo{author}{Nagai, K.},
  \bibinfo{author}{Mizumoto, N.}, \bibinfo{year}{2002}.
\newblock \bibinfo{title}{{Wind tunnel experiments for atmospheric diffusion
  under various stability conditions}}.
\newblock \bibinfo{journal}{J. Jpn. Soc. Atmos. Environ. (in Japanese)}
  \bibinfo{volume}{37}, \bibinfo{pages}{141--154}.
\newblock \URLprefix \url{http://ir.obihiro.ac.jp/dspace/handle/10322/3933}.
\bibitem[{Tomas et~al.(2016)Tomas, Pourquie and Jonker}]{Tomas2016}
\bibinfo{author}{Tomas, J.M.}, \bibinfo{author}{Pourquie, M.J.B.M.},
  \bibinfo{author}{Jonker, H.J.J.}, \bibinfo{year}{2016}.
\newblock \bibinfo{title}{{Stable stratification effects on flow and pollutant
  dispersion in boundary layers entering a generic urban environement}}.
\newblock \bibinfo{journal}{Boundary-Layer Meteorology} \bibinfo{volume}{159},
  \bibinfo{pages}{221--239}.
\bibitem[{Uehara et~al.(2000)Uehara, Murakami, Oikawa and
  Wakamatsu}]{Uehara2000}
\bibinfo{author}{Uehara, K.}, \bibinfo{author}{Murakami, S.},
  \bibinfo{author}{Oikawa, S.}, \bibinfo{author}{Wakamatsu, S.},
  \bibinfo{year}{2000}.
\newblock \bibinfo{title}{{Wind tunnel experiments on how thermal
  stratification affects flow in and above urban street canyons}}.
\newblock \bibinfo{journal}{Atmospheric Environment} \bibinfo{volume}{34},
  \bibinfo{pages}{1553--1562}.
\newblock \DOIprefix\doi{10.1016/S1352-2310(99)00410-0}.
\bibitem[{Wilczak and Phillips(1986)}]{Wilczak1986}
\bibinfo{author}{Wilczak, J.M.}, \bibinfo{author}{Phillips, M.S.},
  \bibinfo{year}{1986}.
\newblock \bibinfo{title}{{An indirect Estimation of Convective Boundary Layer
  Structure for Use in Pollution Dispersion models}}.
\newblock \bibinfo{journal}{Journal of climate and applied meteorology}
  \bibinfo{volume}{25}, \bibinfo{pages}{1609--1624}.
\newblock \DOIprefix\doi{10.1175/1520-0450(1986)025<1609:AIEOCB>2.0.CO;2}.
\bibitem[{Wood et~al.(2010)Wood, Lacser, Barlow, Padhra, Belcher, Nemitz,
  Helfter, Famulari and Grimmond}]{Wood2010}
\bibinfo{author}{Wood, C.R.}, \bibinfo{author}{Lacser, A.},
  \bibinfo{author}{Barlow, J.F.}, \bibinfo{author}{Padhra, A.},
  \bibinfo{author}{Belcher, S.E.}, \bibinfo{author}{Nemitz, E.},
  \bibinfo{author}{Helfter, C.}, \bibinfo{author}{Famulari, D.},
  \bibinfo{author}{Grimmond, C.S.B.}, \bibinfo{year}{2010}.
\newblock \bibinfo{title}{{Turbulent Flow at 190 m Height Above London During
  2006-2008: A Climatology and the Applicability of Similarity Theory}}.
\newblock \bibinfo{journal}{Boundary-Layer Meteorology} \bibinfo{volume}{137},
  \bibinfo{pages}{77--96}.
\newblock \DOIprefix\doi{10.1007/s10546-010-9516-x}.
\bibitem[{Xie et~al.(2013)Xie, Hayden and Wood}]{Xie2013}
\bibinfo{author}{Xie, Z.T.}, \bibinfo{author}{Hayden, P.},
  \bibinfo{author}{Wood, C.R.}, \bibinfo{year}{2013}.
\newblock \bibinfo{title}{{Large-eddy simulation of approaching-flow
  stratification on dispersion over arrays of buildings}}.
\newblock \bibinfo{journal}{Atmospheric Environment} \bibinfo{volume}{71},
  \bibinfo{pages}{64--74}.
\newblock \URLprefix \url{http://dx.doi.org/10.1016/j.atmosenv.2013.01.054},
  \DOIprefix\doi{10.1016/j.atmosenv.2013.01.054}.

\end{thebibliography}


\end{document}